\documentclass[twocolumn, twocolappendix]{aastex631}
\usepackage{gensymb}
\usepackage{amsmath}

\allowdisplaybreaks

\received{May 26, 2025}
\revised{Aug 22, 2021}
\accepted{Sep 5, 2025}

\submitjournal{ApJ}

\graphicspath{{./}{figures/}}

\begin{document}

\title{Neutral gas phase distribution from HI morphology: phase separation with scattering spectra and variational autoencoders}

\author[0000-0002-2679-4609]{Minjie Lei}
\affiliation{Department of Physics, Stanford University, Stanford, CA 94305, USA}
\affiliation{Kavli Institute for Particle Astrophysics \& Cosmology, P.O. Box 2450, Stanford University, Stanford, CA 94305, USA}

\author[0000-0002-7633-3376]{S. E. Clark}
\affiliation{Department of Physics, Stanford University, Stanford, CA 94305, USA}
\affiliation{Kavli Institute for Particle Astrophysics \& Cosmology, P.O. Box 2450, Stanford University, Stanford, CA 94305, USA}

\author[0000-0001-8739-7757]{Rudy Morel}
\affiliation{Center for Computational Mathematics, Flatiron Institute, 162 5th Avenue, New York NY 10010, USA}

\author[0000-0003-3755-7593]{E. Allys}
\affiliation{Laboratoire de Physique de l’Ecole normale supérieure, ENS, Université PSL, CNRS, Sorbonne Université, Université Paris-Cité, Paris, France}

\author[0000-0003-1257-5007]{Iryna S. Butsky}\thanks{NASA Hubble Fellow}
\affiliation{Department of Physics, Stanford University, Stanford, CA 94305, USA}
\affiliation{Kavli Institute for Particle Astrophysics \& Cosmology, P.O. Box 2450, Stanford University, Stanford, CA 94305, USA}

\author[0000-0002-9961-2984]{Caleb Redshaw}
\affiliation{Department of Mechanical Engineering, Stanford University, Stanford, CA 94305, USA}

\author[0000-0003-3806-8548]{Drummond B. Fielding}
\affiliation{Department of Physics,New York University, New York, USA}



\begin{abstract}

Unraveling the multi-phase structure of the diffuse interstellar medium (ISM) as traced by neutral hydrogen (\ion{H}{1}) is essential to understanding the lifecycle of the Milky Way. However, \ion{H}{1} phase separation is a challenging and under-constrained problem. The neutral gas phase distribution is often inferred from the spectral line structure of \ion{H}{1} emission. In this work, we develop a data-driven phase separation method that extracts \ion{H}{1} phase structure solely from the spatial morphology of \ion{H}{1} emission intensity structures. We combine scattering spectra (SS) statistics with a Gaussian-mixture variational autoencoder (VAE) model to: 1. derive an interpretable statistical model of different \ion{H}{1} phases from their multi-scale morphological structures; 2. use this model to decompose the 2D channel maps of GALFA-\ion{H}{1} emission in diffuse high latitude ($|b|>30$\degree) regions over narrow velocity channels ($\Delta v=3$ km/s) into cold neutral medium (CNM), warm neutral medium (WNM), and noise components. We integrate our CNM map over velocity channels to compare it to an existing map produced by a spectrum-based method, and find that the two maps are highly correlated, while ours recovers more spatially coherent structures at small scales. Our work illustrates and quantifies a clear physical connection between the \ion{H}{1} morphology and \ion{H}{1} phase structure, and unlocks a new avenue for improving future phase separation techniques by making use of both \ion{H}{1} spectral and spatial information to decompose \ion{H}{1} in 3D position-position-velocity (PPV) space. These results are consistent with a physical picture where processes that drive \ion{H}{1} phase transitions also shape the morphology of \ion{H}{1} gas, imprinting a sparse, filamentary CNM that forms out of a diffuse, extended WNM.

\end{abstract}


\section{Introduction} \label{sec:intro}

Neutral atomic hydrogen (\ion{H}{1}), as traced by the 21cm hyperfine transition line, is a ubiquitous and critical component of the interstellar medium (ISM). The neutral ISM as traced by \ion{H}{1} emission is composed of the cold neutral medium (CNM), the cold, progenitor material from which star-forming molecular clouds develops; the warm neutral medium, the diffuse gas that extends to the disk-halo interface of galaxies; and the unstable medium, the transitory state between cold and warm phase \citep[see recent review: ][]{mg23-hi}. Resolving the underlying phase structure \ion{H}{1} emission observations, which contain contributions from all \ion{H}{1} phases, is therefore essential to understanding the transfer of matter and energy in galaxies across a wide range of environments and scales.

However, direct observational determinations of \ion{H}{1} phases are limited by the availability of \ion{H}{1} absorption measurements, which are needed to constrain \ion{H}{1} optical depth and derive temperature and density properties \citep{heiles03-ml, murray18-sp}. While \ion{H}{1} emission lines are ubiquitous along every line of sight (LOS) \citep{kalberla09-hi, winkel16-hi}, \ion{H}{1} absorption is limited by the availability of background continuum emission sources. This fundamentally limits our ability to make spatially-resolved, phase-separated \ion{H}{1} maps. As a result, recent work has focused on the development of methods that can extract phase information from \ion{H}{1} emission data alone. A common approach is decomposing the \ion{H}{1} emission spectra into Gaussian components and determining the properties of each component based on its width and amplitude \citep{takakubo66-nh, mebold72-ot, haud07-gd, kalberla18-po}. However, many sightlines have complex line profiles that cannot be uniquely fit by Gaussian components, further complicated by factors such as velocity blending and systematics. More sophisticated techniques have been developed to mitigate these issues including regularization and automated component selection \citep{marchal19-rs, riener20-ag}. Still, the limitations of Gaussian decomposition have prompted the active development of alternative approaches, such as Fourier transform thresholding \citep{marchal24-ft}. Advances in realistic multi-phase ISM simulations have also enabled the application of supervised learning approaches, where neural network models are trained on simulated \ion{H}{1} spectra, and then used to predict the mass fraction of CNM ($f_\mathrm{CNM}$) for a given observed \ion{H}{1} emission sightline \citep{Murray2020-uf, nguyen25-uf}.  However, differences between the synthetic and observed \ion{H}{1} spectra can contribute to bias in the model predictions. 

In short, inferring \ion{H}{1} phase structure from \ion{H}{1} emission data is an important problem that has prompted active development of many innovative methods that furthered our understanding of how spectral line information connects to \ion{H}{1} gas properties. However, in the absence of spatially-resolved absorption line measurements, particularly towards diffuse regions, \ion{H}{1} phase separation remains a challenging, under-constrained problem. In this work, we propose a new data-driven, morphology-based component separation method that complements and advances existing approaches in two important ways: by utilizing spatial structure of \ion{H}{1} emission, and learning phase separation directly from data without relying on simulation training. 

Previous phase-separation methods use only the line-of-sight velocity structure from PPV \ion{H}{1} emission data, with some work making limited use of spatial information to ensure coherence of derived structures \citep{marchal19-rs}. However, there is significant information about \ion{H}{1} phase structure encoded in the spatial dimension of \ion{H}{1} emission as well \citep{lei23-st}. Large-scale, high spatial-resolution \ion{H}{1} emission observations and simulations have both pointed to a general picture of a dense, sparse CNM component, compared to a diffuse, more extended WNM component \citep{kim14-si, Peek2017-ql, fielding23, mg23-hi}. Moreover, studies of \ion{H}{1} emission intensity structures have revealed the presence of magnetically-aligned, filamentary \ion{H}{1} fiber features that are preferentially associated with the CNM \citep{clark14-ma, clark15-fm, kalberla16-hm, kalberla18-po, peek19-cl, clark19_pn}. In general, the magnetized turbulence that shapes the morphology of \ion{H}{1} emission also affects the formation of the CNM out of the WNM \citep{villagran17-cm, das24-cm, munoz25-cm}. Quantifying the distinct morphology features encoded in each \ion{H}{1} phase is an important step to better understand the physical processes that shape the multiphase structure of the ISM. Motivated by these results, in an earlier work we developed a flexible quantification of \ion{H}{1} emission morphology using scattering transform (ST) statistics \citep{allys19-rw, cheng2021-si}, and demonstrated that morphology measures corresponding to small-scale filamentary vs. diffuse features are highly correlated and anti-correlated, respectively, with the mass fraction of CNM \citep{lei23-st}. 

In this work, we extend morphology characterization with ST family of statistics into a full data-driven, morphology-based component separation framework by combining it with a variational autoencoder (VAE) network, and apply the model to \ion{H}{1} phase separation in PPV space. This approach is inspired by the application of VAE models to the separation of seismic signals on Mars \citep{siahkoohi23-ma, siahkoohi24-va}. While the supervised learning approaches adopted in previous work \citep{Murray2020-uf, nguyen25-uf} rely on training with simulations, the unsupervised clustering of VAE networks allows us to identify features that separate different components directly from the observational data, and build a statistical model of each phase that captures realistic multi-scale non-Gaussian information. Moreover, instead of predicting a single value of CNM mass fraction for every sightline, we use the statistical model learned by the VAE to decompose \ion{H}{1} emission maps into CNM, WNM and noise components in narrow velocity channels --- thereby inferring 3D (position-position-velocity) cubes of each component. 

The rest of the paper is organized as follows. In Section \ref{sec:data}, we describe the simulated and observed \ion{H}{1} emission datasets used in this study. Section \ref{sec:method} introduces the ST statistics and VAE network that comprise the component separation model. In Section \ref{sec:results}, we discuss the application and validation of the morphology-based component separation approach with both synthetic and observed \ion{H}{1} emission data. Finally, we discuss the applications and limitations of the proposed method in Section \ref{sec:discuss}, followed by conclusions in Section \ref{sec:conclusion}. Further details of the model and training processes can be found in Appendix \ref{appx:training}. 

\begin{figure*}[t]
\centering
\includegraphics[width=0.9\textwidth]{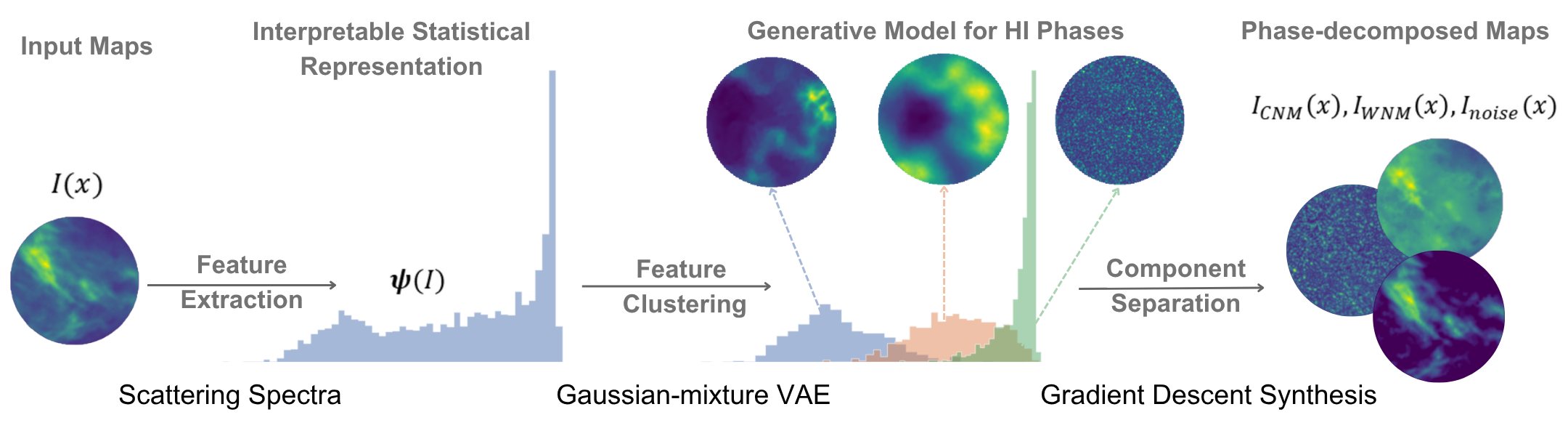}
\caption{A flow chart describing the data-driven, morphology-based phase separation framework. The method is composed of a feature extraction step where we compute the SS statistics of \ion{H}{1} emission channel maps $I(x)$, reducing the input into a more compact but still interpretable representation; a feature clustering step utilizing Gaussian-mixture VAE learn a component-by-component statistical model of \ion{H}{1} phases from the SS statistics; and finally a component separation step where VAE outputs are used as priors to inform phase separation via gradient descent synthesis. }
\label{fig:schematic}
\end{figure*}

\section{Data} \label{sec:data}

In the following section, we describe the observational and simulation data used to test and validate our model. 

\subsection{\ion{H}{1} Emission Data} \label{subsec:galfa}

For observed \ion{H}{1} emission data tracing the local ISM, we use the Data Release 2 of the Galactic Arecibo $L$-band Feed Array Survey \citep[GALFA-\ion{H}{1};][]{Peek2017-ql}. GALFA-\ion{H}{1} maps $\sim32\%$ of the sky, from decl. $-1\degree17\arcmin$ to decl. $+37\degree57\arcmin$ across all R.A., at the highest angular (4\arcmin) and spectral (0.184 km/s) resolution of any large-area Galactic 21 cm emission survey to date. For this study, we restrict our analysis to the high-latitude sky with $|b|>30\degree$ to avoid complex, dense structures at low latitudes with significant saturation and self-absorption. We further restrict the velocity channels we examined to $|v|< 40$ km/s, due to the low signal-to-noise ratio and the lack of significant absorption detected at higher velocities \citep{Murray2015-yu, Murray2020-uf}. 

Motivated by the results that filamentary structures in narrow velocity channels are preferentially associated with the CNM \citep{clark14-ma, clark19_pn, peek19-cl, kalberla20-hm}, we process the GALFA-\ion{H}{1} emission data along the velocity dimension into narrow 3 km/s channel width over the range $|v|< 40$ km/s. The result is a 3D position-position-velocity (PPV) data cube with 4\arcmin\ angular resolution and 3 km/s velocity binning. At each velocity channel, we further divide the sky into $256\times256$ pixel regions (1 pixel = 1\arcmin) that overlap by 50\% of the side length, across 25 velocity channels for a total of 36075 image patches.  Each 2D image patch is treated independently across different velocity channels in all subsequent analysis so that only spatial morphology information is used. 

\subsection{Map of the Cold Neutral Medium} \label{subsec:fcnm}

To compare our phase separation performance with existing methods, we use the cold neutral medium maps produced using a convolutional neural network (CNN) model from \citet{Murray2020-uf} (hereafter M20). The M20 model is trained and tested on augmented, synthetic \ion{H}{1} emission spectra from 3D MHD simulations of the Galactic ISM \citep{kim13-si, kim14-si}. It is then validated against CNM measurements derived using available \ion{H}{1} absorption observations from 21-SPONGE \citep{murray18-sp} and the Millennium survey \citep{heiles03-ml}. Along a specific sightline, the spectral distribution is used to predict a single value representing the CNM mass as a fraction of the total column. We make use of the CNM map produced by applying the M20 model to GALFA-\ion{H}{1} data at high Galactic latitudes ($|b|>30\degree$).

\subsection{3D Multiphase ISM Simulation} \label{subsec:plasmoid}

To validate the performance of the SS+VAE model and complement the studies with \ion{H}{1} emission observation, we apply the method to perform phase separation on synthetic \ion{H}{1} emission maps constructed from a state-of-the-art multiphase ISM simulation from \citet{fielding23} (hereafter F23). F23 is an extremely high resolution ($2048\times2048\times2048$) turbulent magnetohydrodynamic (MHD) simulation with ISM-like heating and cooling functions. Assuming Milky way ISM conditions, including average pressure $P_0\approx10^{3.5}k_B\ \mathrm{K\ cm^{-3}}$, the unstable equilibrium temperature $T_0\approx10^3$ K, and cooling function $\Lambda(T_0)\approx8\times10^{-27}$ \citep{Koyama02-is, jenkins11-is}, the spatial extent of the simulation box translates to a physical spatial scale of $L \sim 75$ pc and a resolution of $\Delta x = L/2048 \sim 0.04$ pc. From the 3D simulation box, we construct synthetic \ion{H}{1} observations following the procedures outlined in \citet{kim14-si}, and map \ion{H}{1} properties such as temperature, density, and velocity from position-position-position (PPP) to PPV space. This results in absorption-corrected synthetic \ion{H}{1} emission patches over velocity channels $|v|<20$ km/s in channels of width of 1 km/s. 

We further process the map at each velocity channel into $128\times128$ pixel patches at a step size of 64 pixels so that every patch partially overlaps. We choose a smaller window size for the synthetic \ion{H}{1} map compared to the GALFA-\ion{H}{1} map due to the higher resolution of the simulation. This leads to 961 patches for each of the 41 velocity channels, for a total of 39401 image patches. Gaussian noise realizations of mean zero and standard deviation of 0.02 K are added to each patch so that we have a similar level of SNR in high-velocity channels as the GALFA-\ion{H}{1} data.  

\section{Methods} \label{sec:method}

Our phase separation approach is composed of three main stages: morphological feature extraction with SS statistics; feature clustering and component identification with a Gaussian-mixture VAE; and phase separation with VAE priors (see Figure \ref{fig:schematic} for a schematic diagram). Here we describe the methodology of each process in detail. 

\subsection{Characterize \ion{H}{1} Morphology with Scattering Spectra} \label{subsec:wss}

Our morphology-based phase separation approach requires a set of summary statistics that can encode morphology information of input fields into a compact, interpretable representation. In this work, we employ the scattering spectra (SS) introduced in \citet{cheng23-ss}. The SS extends the scattering transform statistics described in \citet{mallat2012-gi} \citep[see also][]{cheng2021-si}, and is based on the covariances of ST coefficients. Hereafter we will use ST to denote the general family of scattering transform statistics, and SS to specifically refer to the scattering spectra coefficients. The general ST technique utilizes a set of oriented wavelets as bandpass filters in an iterative process to encode localized, non-Gaussian information from input fields at discrete scales and orientations. The 2D wavelets are constructed via a series of scaling and rotation applied to a mother wavelet $\psi$, in the form of:
\begin{equation}
    \psi_{j,l}(\boldsymbol{x})=2^{-2j}\psi(2^{-j}R_\theta^{-1}\boldsymbol{x})
\end{equation}
The scalings and rotations considered are indexed by $0\leq j\leq J$ and $\theta\in\{l\pi/L,0\leq l\leq L\}$. The scales are sampled in powers of 2 and the maximum number of scales J is set by the dimension of the input field $J=\log_2N-1$. In this work we utilize the Morlet wavelet for $\psi$, which is a sinusoidal wave localized by a Gaussian envelope \citep{morlet82-wl}. 

The standard ST coefficients are computed from the convolutions of wavelets with input fields coupled with a pointwise modulus: $|W_{{j,l}}I(\boldsymbol{x})|=|I(\boldsymbol{x})\star\psi_{j,l}|$. Iterative application of the convolution and modulus operation gives rise to higher order coefficients. The SS statistics extend this definition by computing the covariance between ST coefficients. The sparsity factor $S_1$ and the wavelet power spectrum $S_2$ can be computed from the first order ST coefficients with one wavelet convolution:
\begin{align}
    S_1 (\lambda_1) &= \underset{\boldsymbol{x}}{\mathrm{Ave}}|W_{\lambda_1}I(\boldsymbol{x})| \label{eq:s1} \\ 
    S_2 (\lambda_1) &= \underset{\boldsymbol{x}}{\mathrm{Ave}}\left<W_{\lambda_1}I(\boldsymbol{x}), W_{\lambda_1}I(\boldsymbol{x})\right> = \underset{\boldsymbol{x}}{\mathrm{Ave}}|W_{\lambda_1}I(\boldsymbol{x})|^2 \label{eq:s2}
\end{align}
Here $\lambda$ is the wavelet index in place of $(j, l)$. The higher-order moments $S_3$ and $S_4$ can be computed from the covariances of higher-order ST coefficients, which employ two successive wavelet convolutions and modulus operations: 
\begin{align}
    S_3 (\lambda_1, \lambda_2) &= \underset{\boldsymbol{x}}{\mathrm{Ave}}\left<W_{\lambda_1}I(\boldsymbol{x}), W_{\lambda_1}|W_{\lambda_2}I(\boldsymbol{x})|\right> \label{eq:s3} \\
    S_4 (\lambda_1, \lambda_2, \lambda_3)   &= \underset{\boldsymbol{x}}{\mathrm{Ave}}\left<W_{\lambda_1}|W_{\lambda_2}I(\boldsymbol{x})|, W_{\lambda_1}|W_{\lambda_3}I(\boldsymbol{x})|\right> \label{eq:s4}
\end{align}

Together the wavelet moments define the vector of scattering spectra $\Psi=[S_1(\boldsymbol{x}), S_2(\boldsymbol{x}), S_3(\boldsymbol{x}), S_4(\boldsymbol{x})]$ we will use to describe the morphological features of \ion{H}{1} emission fields. Since wavelet filters sample scales dyadically up to field dimension $N$ and orientations discretely up to $L$, the total number of coefficients for the SS representation is of order $O(log^3N*L^3)$. The additional covariance information encoded by the SS statistics means that it contains significantly more coefficients than the standard ST statistics, which is of order $O(log^2N*L^2)$. However, an important benefit of the additional information is that SS coefficients are highly descriptive statistics of the input field, such that we can build a generative model under SS constraints to produce more realistic image synthesis \citep{allys20-wph, cheng23-ss}.  This process relies on the microcanonical gradient descent models introduced in \citet{bruna18-mm}. The idea is to generate new realizations of a random field by iteratively transforming Gaussian samples conditioned on a prescribed set of SS statistics through a gradient descent process. In practice, we produce an image realization $I_c$ from a set of target SS statistics $S_c$ by initializing the image as a Gaussian noise realization, and then evolving it through a gradient descent optimization process to minimize the loss criteria:
\begin{equation} \label{eq:loss_syn}
    \mathcal{L}_{\rm syn}(I_c)=\big|\big|\Psi(I_c)-S_c\big|\big|^2_2
\end{equation}
where $c=$ CNM, WNM, or noise. $\Psi(I_c)$ computes the SS statistics of $I_c(\boldsymbol{x})$ as specified in Equations \ref{eq:s1}-\ref{eq:s4}.

We can extract from the full set of coefficients a few compact summary statistics with convenient interpretations. First, if we expect isotropically distributed features from the input fields, we can average over the absolute orientations. While individual \ion{H}{1} emission structures can be highly anisotropic, and this anisotropy is captured by the SS statistics, we do not expect the absolute orientation of the anisotropic structures on the sky to be statistically important. This simplification collapses the $S_1$, $S_2$ coefficients to depend only on scale, and reduces higher-order coefficients to depend only on relative orientations $|l_1-l_2|$ or $|l_1-l_3|$. Then the $S_2$ coefficient is only a function of scale, and analogous to the isotropic power spectrum:
\begin{equation} \label{eq:power}
    \mathrm{Power:\ } S_2(j) = \underset{\boldsymbol{l}}{\mathrm{Ave}}\left(\underset{\boldsymbol{x}}{\mathrm{Ave}}|W_{j,l}I(\boldsymbol{x})|^2\right) 
\end{equation}
The ratio between the first and second order coefficients is a measure of sparsity at different scales:
\begin{equation} \label{eq:sparsity}
    \mathrm{Sparsity:\ } \frac{S_1^2}{S_2}(j)= \frac{\underset{\boldsymbol{l}}{\mathrm{Ave}}\left(\underset{\boldsymbol{x}}{\mathrm{Ave}}|W_{j,l}I(\boldsymbol{x})|\right)^2}{\underset{\boldsymbol{l}}{\mathrm{Ave}}\left(\underset{\boldsymbol{x}}{\mathrm{Ave}}|W_{j,l}I(\boldsymbol{x})|^2 \right)}
\end{equation}
The ratio decreases as the sparsity of the field increases, as more relative power is carried by the second order coefficients. Finally, a measure of feature alignment can be defined from the higher-order coefficients where the orientations of different wavelets are aligned versus anti-aligned. In that case, successive wavelet convolution probes the clustering of features along parallel/perpendicular orientation at multiple scales, indicating aligned/anti-aligned structures. The ratio of these corresponds to a measure of filamentarity in the input image. In the case of the isotropic $S_3$ coefficients, this corresponds to:
\begin{equation} \label{eq:linearity}
    \mathrm{Filamentarity:\ } \frac{S_{3\parallel}}{S_{3\perp}}(j_1,j_2)=\underset{\boldsymbol{l_1}}{\mathrm{Ave}}\ \frac{S_3(j_1,j_2,l_1,l_2=l_1)}{S_3(j_1,j_2,l_1,l_2\perp l_1)}
\end{equation}
We utilize the full set of isotropic SS coefficients when training the VAE model and to perform phase separation, but use the more compact and interpretable summary statistics (Equations \ref{eq:power}-\ref{eq:linearity}) when interpreting and comparing results.  

When deriving the isotropic SS coefficients from the 2D image patches described in Section \ref{subsec:galfa}
and \ref{subsec:plasmoid}, we choose the number of scales based on patch size $J=\log_2N-1$ and number of orientations to be $L=4$. For the $256\times256$ pixel size GALFA-\ion{H}{1} patches, this corresponds to $J=7$ and a total of 1470 SS coefficients per patch, while the $128\times128$ synthetic \ion{H}{1} patches translates to $J=6$ and a total of 992 coefficients per patch. Additionally, to mitigate non-periodic boundary effects when computing SS statistics on square image patches, we apply a circular apodization mask tapered with a cosine function to each patch, with the apodization scale set to the patch size. The final dimensions of the SS representation of the input image patches are $(N_{\rm \# patches}, K_{\rm \#coefficients})=(36075, 1470)$ and $(N_{\rm \# patches}, K_{\rm \#coefficients})=(39401, 992)$ for the GALFA-\ion{H}{1} and F23 simulation datasets respectively. 

\subsection{Feature Clustering with Gaussian-Mixture Variational Autoencoder} \label{subsec:vae}

The second step of our data-driven method is a clustering algorithm that can effectively learn a component-by-component model of the input data, which is performed in the feature space (the space of the SS coefficients) introduced in the previous section.

In a simplified setting, one might assume that the distribution of the SS statistics for a single (hidden) component follows a Gaussian distribution. Under this assumption, it can be natural to fit a Gaussian Mixture Model (GMM)~\citep{xu1996convergence}, using the expected number of components as the number of Gaussians and fitting the weights, means, and variances of each Gaussian from the data.

However, the distribution of SS statistics for a given component can hardly be assumed to be Gaussian. There is no guarantee that perturbing the SS in one direction is as likely as perturbing it in the opposite direction — a symmetry that underlies the Gaussian assumption (i.e., zero skewness). Nevertheless, visualizations of SS coefficients~\citep{cheng23-ss} reveal rigid patterns — such as regular decay along angles or scales — which constrain the set of admissible statistics. Just as not every power spectrum corresponds to a physically plausible process, this suggests that SS may be effectively approximated using a lower-dimensional representation.

A Gaussian mixture VAE model~\citep{digma13-va, dilo16-gm} precisely addresses these two challenges by learning a Gaussian mixture model, not in the original SS space, but in a learned latent space of reduced dimensionality. VAEs are composed of an encoder that compresses data into a latent space, and a decoder that reconstructs data from latent representations. Here, the encoder is used to map the SS coefficient representation of our input \ion{H}{1} emission fields into a low-dimensional Gaussian mixture distribution over the latent space, enabling the clustering of features into different Gaussian mixture components. Then the decoder provides a generative model that maps each Gaussian mixture latent distribution back to an SS coefficient distribution for each identified component. 

Suppose $s$ denotes the input SS coefficients, and $y$, $z$ are the categorical latent and Gaussian mixture variables respectively. The encoder is an inference network that parametrizes the approximate posterior $q_{\phi}(z,y|s)=q_{\phi}(z|y,s)q_{\phi}(y|s)$ with parameters $\phi$ as:
\begin{align} \label{eq:encoder} \nonumber
    q_{\phi}(y|s) &= \mathrm{Cat}(\boldsymbol{\pi}(s;\phi)) \\ 
    q_{\phi}(z|y,s) &= \mathcal{N}\big(z|\boldsymbol{\mu}_z(s, y;\phi), \rm{diag}(\boldsymbol{\sigma}^2_z(s,y;\phi))\big)
\end{align}
where $\mathrm{Cat}$ denotes categorical distribution. $\boldsymbol{\pi}(s;\phi)$ represents the cluster membership probability for each input $s$, and is parametrized by a deep net. $q_{\phi}(z|y,s)$ is modeled as a Gaussian distribution with mean $\boldsymbol{\mu}_z(s, y;\phi)$ and diagonal covariance $\boldsymbol{\sigma}^2_z(s,y;\phi))$. 

On the other hand, the decoder is a generative network that parametrizes the joint distribution $p_\theta(s, y, z
)=p_\theta(x|z)p_\theta(z|y)p(y)$ with parameters $\theta$ as:
\begin{align} \label{eq:decoder} \nonumber
    p_\theta(y) &= \mathrm{Cat}(\mathbb{I}_c/c) \\ 
    p_\theta(z|y) &= \mathcal{N}\big(z|\boldsymbol{\mu}_z( y;\phi), \boldsymbol{\sigma}^2_z(y;\phi)\big) \\ \nonumber
    p_\theta(s|z) &= \mathcal{N}\big(s|\boldsymbol{\mu}_s( z;\phi), \boldsymbol{\sigma}^2_s(z;\phi)\big)
\end{align}
where $c$ denotes the number of components in the Gaussian mixture distribution, and $\mathbb{I}_c$ is the identity matrix of dimension $c\times c$. $p_\theta(z|y)$ and $p_\theta(s|z)$ are chosen to be Gaussian distributions with their respective mean and diagonal covariances. 

The training objective for a VAE model is to minimize the difference between the parametrized and true joint distribution, as measured by the Kullback-Leibler ($\mathbb{KL}$) divergence. The evaluation of the likelihood is intractable in practice, therefore it is approximated by the evidence lower bound (ELBO), leading to the following loss function:
\begin{align} \label{eq:vae_loss} \nonumber
    \underset{\theta,\phi}{\rm{min}}\underset{s\sim p(s)}{\mathbb{E}}&\bigg[\underset{z\sim q_\phi(z|y,s)}{\mathbb{E}}\big[\underbrace{-\log p_\theta(s|z)}_{\rm{Reconstruction\ loss}}\big]+\underbrace{\mathbb{KL}\big(q_\phi(y|s)||p_\theta(y)\big)}_{\rm{Categorical\ prior\ loss}} \\
    &+\underset{y\sim q_\phi(y|s)}{\mathbb{E}}\underbrace{\mathbb{KL}\big(q_\phi(z|y, s)||p_\theta(z|y)\big)}_{\rm{Gaussian\ prior\ loss}}\bigg]
\end{align}

The expectations $\mathbb{E}$ in the loss terms can be approximated using Monte Carlo integration over distribution samples. The architecture and training details of the Gaussian mixture VAE model used for this study are given in Appendix \ref{appx:training}. 

The result of applying the Gaussian VAE model to input \ion{H}{1} emission data is a set of statistical models for the distribution of the CNM, WNM, and noise components. We can then sample from the VAE distributions to obtain a set of SS statistics $S^i_c=S_{\rm CNM}^i, S_{\rm WNM}^i, S_{\rm noise}^i$ that describe each component, where $i=1, ..., M$ iterates over the total number of samples. We further construct a set of synthesized images $\tilde{I}^i_c=\tilde{I}_{\rm CNM}^i, \tilde{I}_{\rm WNM}^i, \tilde{I}_{\rm noise}^i$ from $S^i_c$ through the gradient descent process described by \ref{eq:loss_syn}. We utilize $S^i_c$ and $\tilde{I}_c^i$ as priors to inform phase separation in the next step.

\subsection{Phase Separation with VAE Cluster Priors} \label{subsec:phase_sep}

Having learned a component-separated clustering in the SS coefficient space, we use the VAE outputs as prior information to perform phase separation. Utilizing ST statistics as constraints to perform statistical component separation has been successfully applied for problems such as denoising and cosmic infrared background separation \citep{blancard21-dn, delouis22-st, auclair24-st}. In this study, our goal is to decompose input \ion{H}{1} emission fields into separate phases plus a noise field $I(\boldsymbol{x})=I_{\rm CNM}(\boldsymbol{x})+I_{\rm WNM}(\boldsymbol{x})+I_{\rm noise}(\boldsymbol{x})$. Here, we use WNM to denote any physical component of the \ion{H}{1} emission that is not CNM, i.e., ``WNM" represents a mixture of WNM and UNM. We discuss the possibility of separating the UNM component in more detail when examining the results of applying the VAE model in Section \ref{sec:results}. 

There are many different ways to utilize the VAE priors to perform phase separation, including machine learning algorithms like moment networks \citep{jeffrey22-nn}. Here, we approach decomposing $I(\boldsymbol{x})$ through a simple gradient descent process similar to what is described by Equation \ref{eq:loss_syn}, incorporating $I_{\rm CNM}^i, I_{\rm WNM}^i, I_{\rm noise}^i$ as prior information. We first perform a denoising step to detect and separate the noise component from the input images, and then a phase separation step to separate the WNM from the denoised image, leaving the residual image as the CNM component. The order of the component separation is motivated by the fact that the high-latitude diffuse gas examined in this study is dominated by warm gas. The gradient descent loss terms for the denoising step are given by:
\begin{align} \label{eq:loss_denoise} \nonumber
    \mathcal{L}_{\rm prior}(I_{\rm noise}) &= \sum_{i=1}^{M}\frac{\big|\big|\Psi(I_{\rm noise})-\Psi(\tilde{I}_{\rm noise}^i)\big|\big|^2_2}{\sigma^2_i\big(\Psi(\tilde{I}_{\rm noise}^i)\big)}\\
    \mathcal{L}_{\rm cross}(I_{\rm noise}) &= \sum_{i=1}^{M}\frac{\big|\big|\Psi(I_{\rm noise},I-I_{\rm noise})\big|\big|^2_2}{\sigma^2_i\big(\Psi(\tilde{I}_{\rm noise}, I)\big)} \\ \nonumber
    \mathcal{L}_{\rm reg}(I_{\rm noise}) &=w_{\rm reg}\sum_{n=1}^{M}\frac{\big|\big|\Psi(I-I_{\rm noise})-\Psi(\tilde{I}_{\rm WNM}^i)\big|\big|^2_2}{\sigma^2_i\big(\Psi(\tilde{I}_{\rm WNM}^i)\big)}
\end{align}
The prior loss term ($\mathcal{L}_{\rm prior}$) enforces consistency between the statistics of the noise component and the VAE noise prior synthesis. The cross loss term ($\mathcal{L}_{\rm cross}$) enforces statistical independence between the noise and the residual data component. We normalize each term by its corresponding variance over number of VAE samples, such that the loss converges to $\mathcal{O}(1)$.  We also include a weighted regularization term ($\mathcal{L}_{\rm reg}$) that promotes agreement between the residual component and the VAE WNM prior synthesis. While the data residual $I-I_{\rm noise}$ is a mixture of CNM and WNM components, in many low column density regions of the GALFA-\ion{H}{1} high latitude footprint at high velocity channels, there is little to no CNM content. These regions usually also have lower signal-to-noise (SNR) ratio, and we found that including $\mathcal{L}_{\rm reg}$ improves denoising performance. A weighting of $w_{\rm reg}=0.1$ is adopted for input images with SNR $<$ 2, and $w_{\rm reg}=0.01$ is applied to images with higher SNR. We initialize $I-I_{\rm noise}$ to be the result after convolving the input image with a Gaussian kernel with $\sigma=1$. In principle, $I_{\rm noise}$ can be initialized to be zero, but we found that the current initialization results in faster convergence. With this initialization, $I_\mathrm{noise}$ is akin to high-pass filtering, thus isolating the highest-frequency oscillations, which are often noise and artifacts mixed with small-scale data features. We use this as a starting point for the denoising algorithm which then separates the noise from high-frequency data features enforced by the prior and statistical independence loss terms.   

The phase separation step is then performed on the denoised residual $I'(\boldsymbol{x})\equiv I(\boldsymbol{x})-I_{\rm noise}(\boldsymbol{x})$ with the following loss terms:
\begin{align} \label{eq:loss_decompose} \nonumber
    \mathcal{L}_{\rm prior}(I_{\rm WNM}) &= \sum_{i=1}^{M}\frac{\big|\big|\Psi(I_{\rm WNM})-\Psi(\tilde{I}_{\rm WNM}^i)\big|\big|^2_2}{\sigma^2_i\big(\Psi(\tilde{I}_{\rm WNM}^i)\big)}\\
    \mathcal{L}_{\rm reg}(I_{\rm WNM}) &=w_{\rm reg}\bigg|\bigg|\frac{(I'-I_{\rm WNM})}{I'^2}\bigg|\bigg|^2_2
\end{align}
The prior loss term here similarly ensures agreement between the WNM component and the VAE WNM prior synthesis. We do not include a cross term between $I_{\rm WNM}$ and $I_{\rm CNM}=I-I_{\rm WNM}$ since we do expect CNM and WNM in the same region to be statistically correlated, as CNM forms out of WNM gas as a result of thermal instability \citep{field69-ph, wolfire03-hi, mg23-hi}. Finally, a weighted regularization term is applied to penalize high CNM fractions for the lowest column density pixels in every $4\degree \times4\degree$ image patch, which are more likely to be noise dominated. We adopt a weighting factor $w_{\rm reg} = 1$ for the phase separation step. $I_{\rm WNM}$ is initialized to be the original input image $I'$, so that $I_{\rm CNM}$ is initialized to be zero. Both $I_{\rm WNM}$ and $I_{\rm CNM}$ are enforced to have positive pixel values. 

Since we choose to perform phase separation in steps where components are separated from the input one by one using their respective priors, the CNM priors are not directly used in the end. The motivations are twofold. First, utilizing CNM and WNM prior constraints simultaneously would require us to fully characterize the statistical correlation between CNM and WNM. Currently the VAE clustering provides prior information for the statistical distribution of each component, but not the cross correlation between components. Future improvement to the model can potentially be made by incorporating cross-SS coefficients to characterize cross-correlation information between input fields \citep{cheng23-ss}. Secondly, the \ion{H}{1} emission in our target diffuse high latitude region is generally more WNM-dominated \citep{murray18-sp, Murray2020-uf}. Therefore, there is a fuller range of WNM distribution represented in the input data, while there are few input patches with high CNM. As a result, the CNM priors are likely an incomplete statistical sampling of the full CNM distribution due to the imbalanced nature of the input dataset, specifically the lack of high CNM patches. This is not a problem for the proposed phase separation as long as we can reliably separate the noise and WNM component from the input data, leaving the residual as the CNM component. 


\begin{figure*}[t]
\centering
\includegraphics[width=0.9\textwidth]{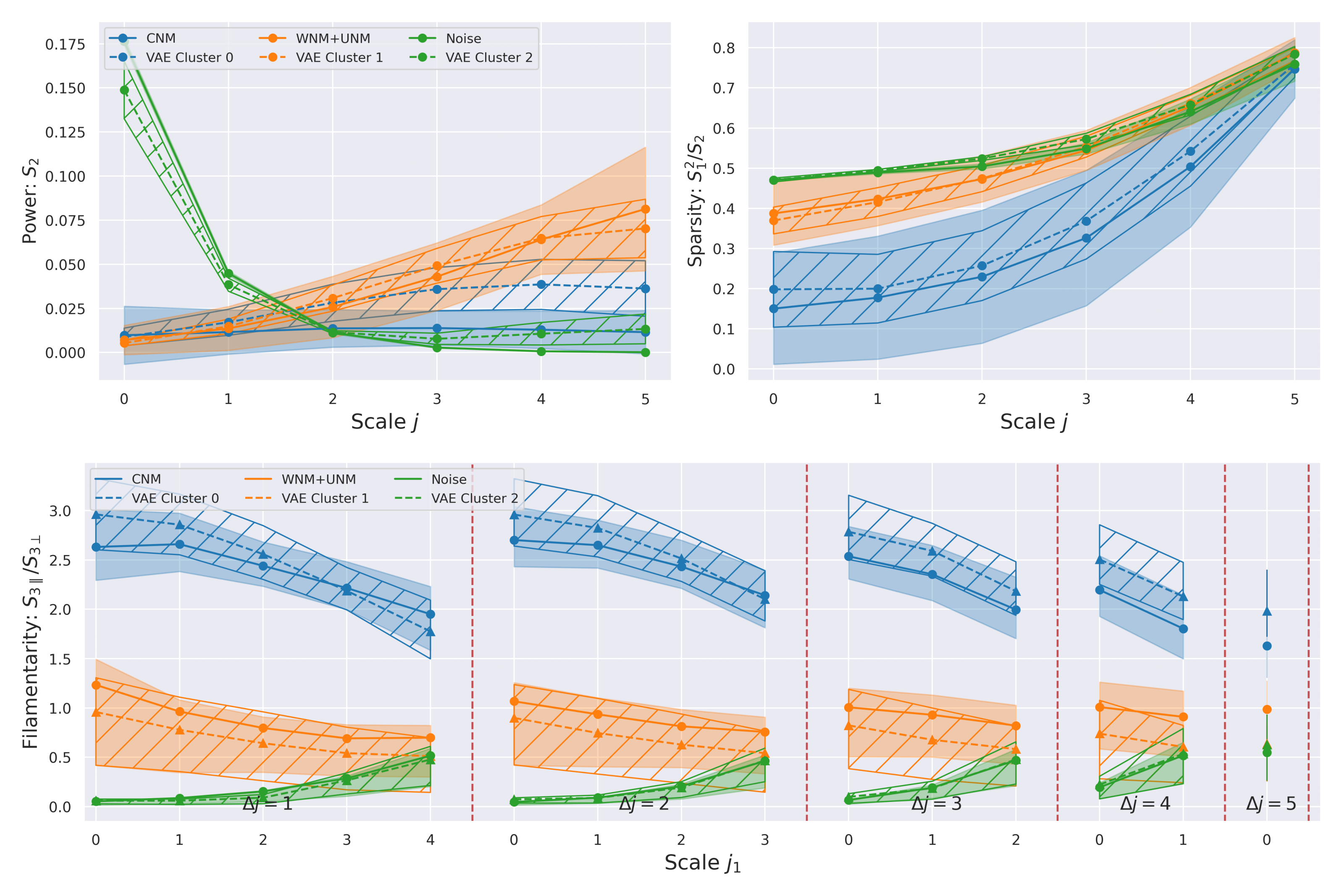}
\caption{Distribution of the SS summary statistics for the identified VAE clusters compared to that of the true phase distribution from the F23 simulation. The top left and right panels correspond to power and sparsity, while the bottom panel correspond to filamentarity, as specified by Equations \ref{eq:power}-\ref{eq:linearity}. The three distinct components identified by the VAE model agree very well in the SS representation with the CNM, WNM+UNM, and noise respectively. }
\label{fig:prior_vs_true_plasmoid}
\end{figure*}

\section{Results} \label{sec:results}

\subsection{Validation with Multiphase ISM Simulation} \label{subsubsec:plasmoid_validation}

To demonstrate the applicability of the proposed unsupervised component separation method to ISM phase separation problems, we first apply the model to the multiphase ISM simulation from \citet{fielding23}. There are several key differences to note between the simulation and the observed GALFA-\ion{H}{1} data. First, assuming Galactic ISM conditions, the $128\times128$ pixel synthetic \ion{H}{1} patches translate to a physical size of $\sim5\times5$ pc with resolution of 0.04 pc. Compared to the GALFA-\ion{H}{1} resolution of 4\arcmin, the F23 simulation describes structures in a much smaller volume at a finer scale. For example, assuming a fiducial distance of 200 pc, the approximate distance to the Local Bubble wall \citep{zucker22-lb}, the GALFA-\ion{H}{1} angular resolution translates to a physical resolution of $\sim0.24$ pc. Moreover, the F23 simulation is, on average, more cold gas dominated and reaches higher column densities than the high latitude diffuse GALFA-\ion{H}{1} region. Therefore, we do not necessarily expect the statistical description of simulated emission components to also describe the GALFA-\ion{H}{1} data, but we can use the simulations to validate our methodology in a case where we know the ground truth.

\begin{figure*}[t]
\centering
\includegraphics[width=0.9\textwidth]{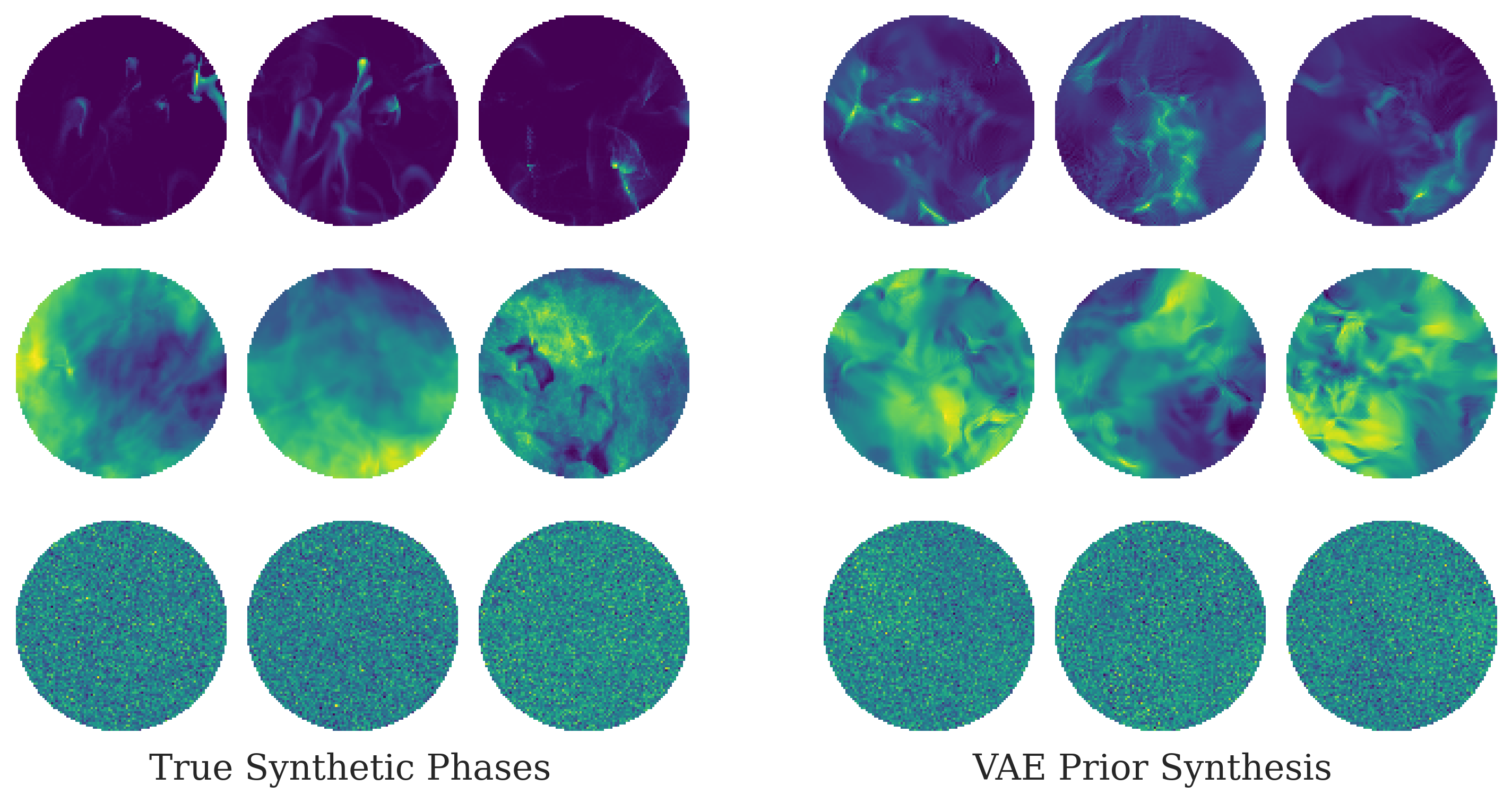}
\caption{Image realizations synthesized from the SS representations identified by the VAE model from the F23 simulation dataset, compared to the ground truth synthetic CNM, WNM+UNM, and noise patches. The left panels are the ground truth synthetic \ion{H}{1} images while the right panels correspond to the SS+VAE synthesis. The top, middle, and bottom rows corresponds to CNM, WNM+UNM, and noise respectively. All images are normalized to have zero mean and unit standard deviation.}
\label{fig:vae_synthsis_plasmoid}
\end{figure*}

\begin{figure*}[t]
\centering
\includegraphics[width=0.9
\textwidth]{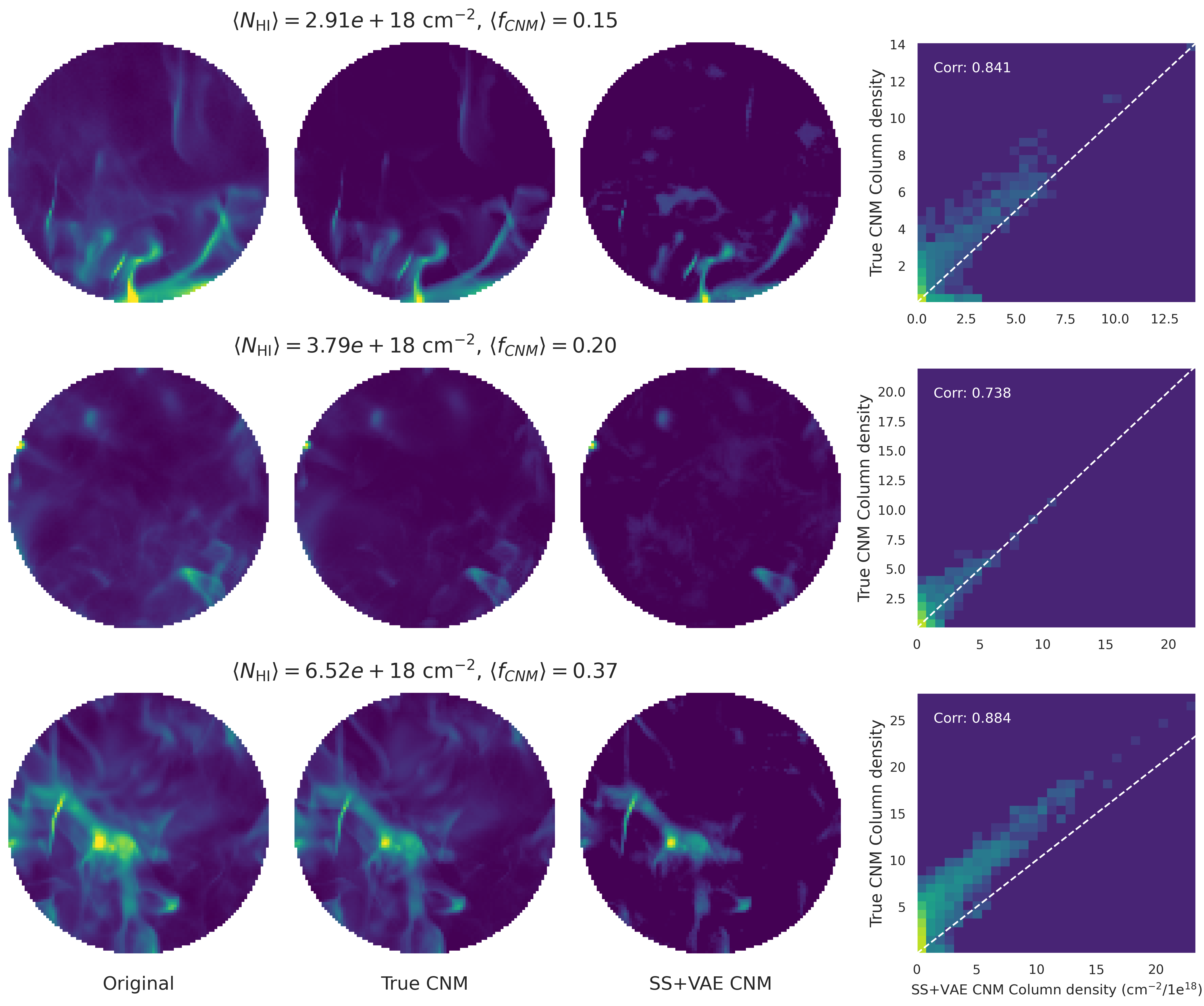}
\caption{Result of SS+VAE CNM separation performed on the F23 simulation dataset compared to the true CNM distribution for a few different velocity channels of a sample sightline. Left: original synthetic \ion{H}{1} emission. Middle: True CNM column density. Right: CNM column density predicted by the SS+VAE model. The last panel shows the pixel-by-pixel correlation between the true and predicted CNM column density, with the white dashed line indicating one-to-one correlation.}
\label{fig:phase_decomp_plasmoid}
\end{figure*}

\subsubsection{Identify Simulated \ion{H}{1} Phases with VAE Clustering} \label{subsubsec:vae_prior_sim}

As described in Section \ref{subsec:wss}, the input SS coefficients of dimension  $(N_{\rm \# patches}, K_{\rm \#coefficients})=(39401, 992)$ are derived from the full sets of $128\times128$ pixel synthetic \ion{H}{1} patches. We then apply the Gaussian-mixture VAE model described in Section \ref{subsec:vae} on the SS input to learn component-separated distributions and identify different phase structures from the synthetic \ion{H}{1} emission maps. Motivated by the physical expectation that \ion{H}{1} emission maps contain CNM, UNM, WNM, and noise components, we choose the number of Gaussian mixture components to be $c=4$. However, when we examine the distribution of our four VAE components in latent space, we find that the synthetic \ion{H}{1} dataset is clustered into only three distinct components, with the fourth overlapping significantly with the third component. In the subsequent discussion, we will compare the properties of the first three distinct components to the true phase distribution of the synthetic \ion{H}{1} emission, and discuss possible explanations for why the VAE can only identify three non-redundant components and how they correspond to the expected \ion{H}{1} phases. Note that we choose to keep 4 clusters in this section to allow a better comparison with the results obtained for observational data.

To compare the distinct VAE clusters with the true phase distribution, we follow \citet{kim14-si} and apply temperature cuts of $T_k<184\ \rm K$, $184\ \rm K<T_k< 5050\ \rm K$, and $T_k> 5050\ \rm K$ respectively for the CNM, UNM and WNM to derive sets of phase-separated patches. We combine the synthetic WNM and UNM emission maps and compare CNM, WNM+UNM, and Gaussian noise to the three distinct VAE components respectively. The comparison is first done in the SS representation: we apply the SS to the phase-separated patches and compare their distribution to the distribution of the VAE clusters. We show the results for the set of summary statistics described in Equations \ref{eq:power}-\ref{eq:sparsity} in Figure \ref{fig:prior_vs_true_plasmoid}. The three distinct clusters identified by the VAE model trained on the synthetic \ion{H}{1} mixture closely match the true distribution of the CNM, WNM+UNM, and noise, respectively. The additional fourth VAE component overlaps significantly with cluster 2, and is therefore not shown in \ref{fig:prior_vs_true_plasmoid}. The summary statistics are consistent with what we expect for the morphological properties of each phase. The CNM is more sparse and filamentary than the WNM+UNM across scales, and particularly at smaller scales, while also having a flatter wavelet power spectrum \citep{lei23-st}. The noise component is the least sparse and filamentary, and has most of its power at small scales. 

In the synthetic maps the VAE identifies the separation of noise and CNM from the WNM+UNM well, but is not able to separate UNM from WNM even when choosing number of VAE components to be $c=4$. This is likely due to the WNM and UNM being much closer to each other in morphology measure space than to the other phases. The L2 distance of the SS mean between the simulated CNM and WNM is 3 times that between the UNM and WNM, while the N-dimensional Wasserstein distance \citep{panaretos19-wa} of the SS distribution is two times higher for CNM/WNM compared to UNM/WNM. Furthermore, the synthetic \ion{H}{1} emission is highly CNM-dominated, and the UNM and WNM together account for only ~20\% of the total mass fraction. As a result, the VAE model trained on the simulated total \ion{H}{1} emission more readily separates the CNM and noise components from a mixture of WNM and UNM.

We can further examine how well the VAE clusters describe the true underlying phases by directly synthesizing images from the SS statistics sampled from each cluster. The process is described in Section \ref{subsec:phase_sep} and Equation \ref{eq:loss_syn}. A few examples of synthesis for each cluster are shown in Figure \ref{fig:vae_synthsis_plasmoid}. The morphological features of the synthesized images are visually similar when compared with the ground truth synthetic CNM, WNM+UNM, and noise patches. 

\subsubsection{Decompose Simulated \ion{H}{1} Emission with VAE Priors} \label{subsubsec:phase_sep_sim}

Having identified distinct components from the simulation dataset and produced synthesized realizations that correspond well to each phase, we use the image synthesis as prior information to perform phase separation on synthetic \ion{H}{1} maps using procedures described in Section \ref{subsec:phase_sep}. The denoising task is trivial in this case since we only have well-behaved Gaussian noise. We will present denoising in more detail when discussing results with the GALFA-\ion{H}{1} data. We perform phase separation on the denoised images to separate CNM from the WNM+UNM using VAE priors with optimization constraints described in Equations \ref{eq:loss_decompose}. The CNM is initialized to be zero, while the WNM+UNM component is initialized as the original synthetic emission map. The components are then evolved according to the statistical distribution of the VAE priors. We show the results for a few example sightlines in Figure \ref{fig:phase_decomp_plasmoid}. Even though the proposed model is a statistical component separation method and does not necessarily recover deterministic structures at all scales, we found that the CNM produced by the SS-VAE model generally agrees well with the true CNM distribution even on a pixel-by-pixel basis. For image patches with an average CNM fraction between 10\% and 50\%, the Pearson correlation coefficient between the ground truth and phase-separated CNM emission maps is in the range of $\sim0.7-0.9$. The agreement becomes worse for patches with much higher CNM fraction, and the predicted CNM fraction generally underestimates the true CNM fraction. These are the regions at high column density where CNM structures fill the whole patch. In these cases, as the CNM takes up the entire region they appear more diffuse morphologically and are outliers to the sparse and filamentary morphology that characterizes most CNM patches as shown in Figure \ref{fig:vae_synthsis_plasmoid}. However, since we restrict our analysis of real GALFA-\ion{H}{1} data to diffuse high latitude regions where the CNM fraction is generally much lower and the scale probed by the observation is much larger than what is represented in the simulation, we do not expect to encounter the same outliers. Therefore, the proposed phase separation method performs well in the regime most relevant to the diffuse ISM. In the following section, we present the result of applying the same phase separation technique to the GALFA-\ion{H}{1} data and compare the results with existing methods.

\subsection{Phase Separation Results with GALFA-\ion{H}{1} Data} \label{subsec:phase_sep_res}

As described in Section \ref{subsec:wss}, we compute SS coefficients from input GALFA-\ion{H}{1} patches to derive a compact morphological representation of dimension $(N_{\rm \# patches}, K_{\rm \#coefficients})=(36075, 1470)$. We then apply the VAE model to identify distinct components from the coefficient representation, and use the identified components as prior information to decompose the original input GALFA-\ion{H}{1} patches to arrive at phase-decomposed maps in PPV space. 

\subsubsection{Identify \ion{H}{1} Phases with VAE Clustering} \label{subsubsec:vae_prior_galfa}


\begin{figure*}[t]
\centering
\includegraphics[width=0.9\textwidth]{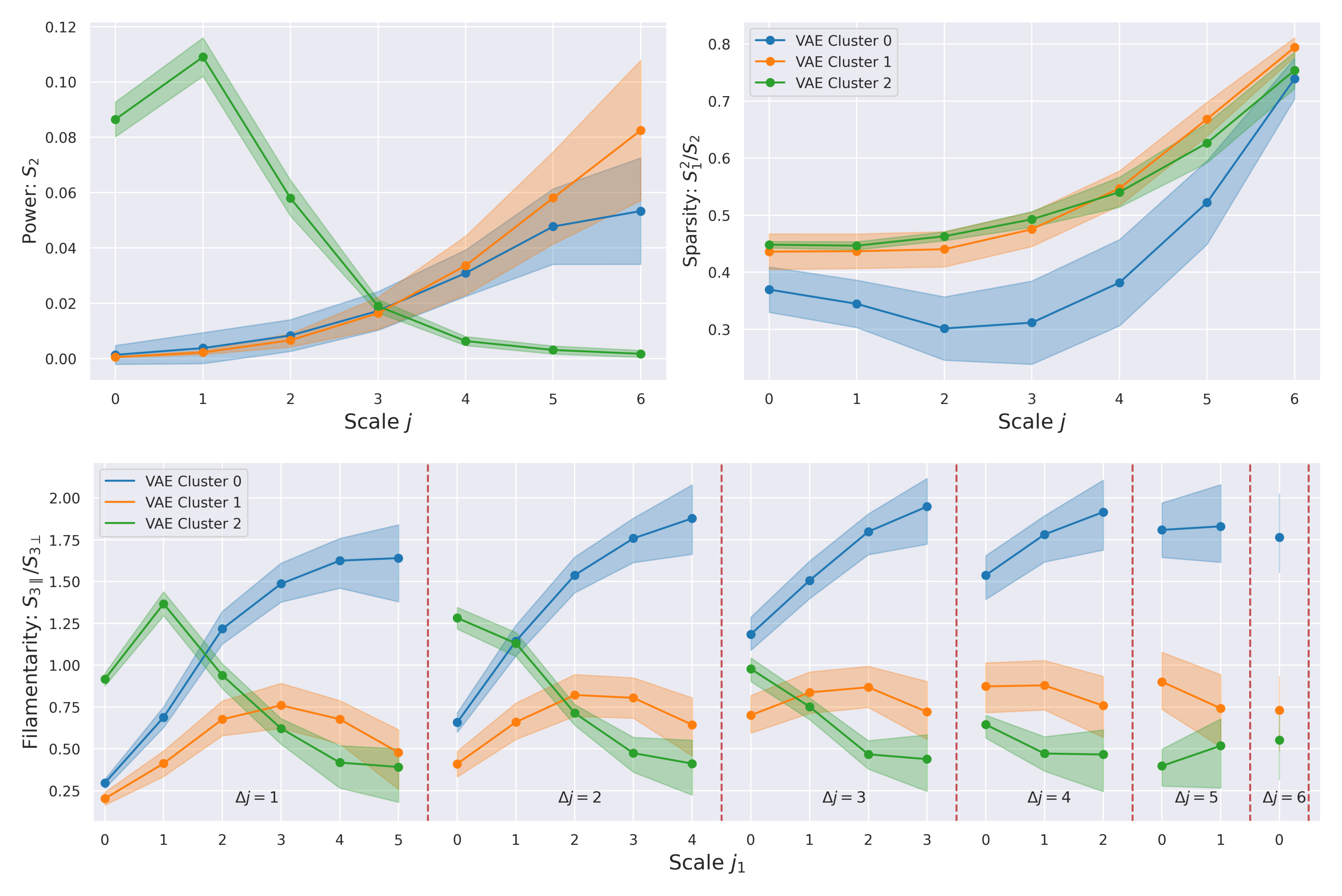}
\caption{Distribution of the SS summary statistics for the distinct clusters identified by the VAE model from the GALFA-\ion{H}{1} dataset. The top left and right panels correspond to power and sparsity, while the bottom panel correspond to filamentarity, as specified by Equations \ref{eq:power}-\ref{eq:linearity}. The morphological interpretations behind each coefficient shows a consistent picture where cluster 0 describes a sparse, filamentary CNM, cluster 1 describes a diffuse WNM, and cluster 2 corresponds to the noise component. The qualitative scale-dependent behavior of each cluster is also consistent with the simulation results in Figure \ref{fig:prior_vs_true_plasmoid}.}
\label{fig:prior_dist_galfa}
\end{figure*}

\begin{figure*}[t]
\centering
\includegraphics[width=0.9\textwidth]{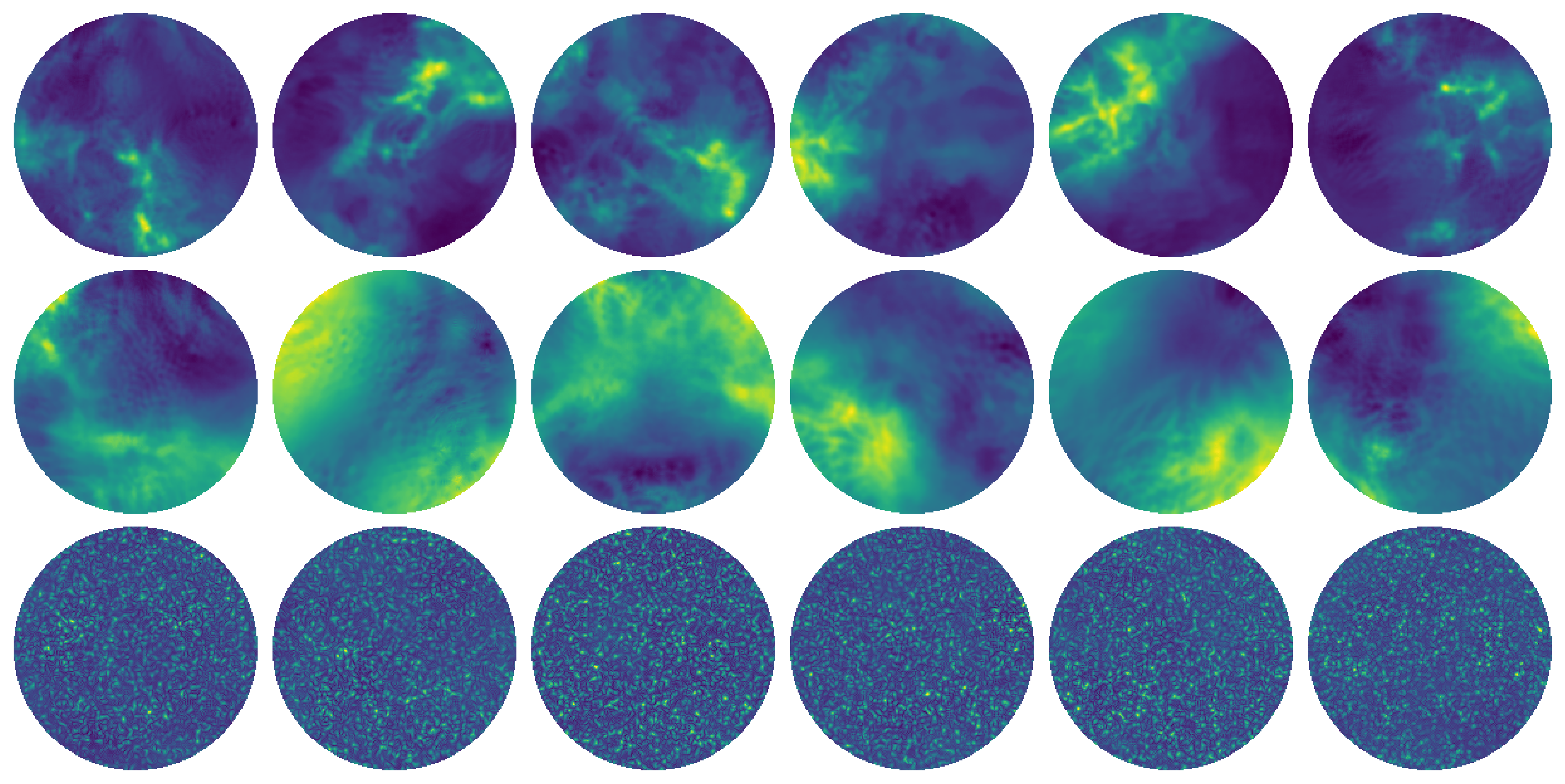}
\caption{Image realizations synthesized from the SS coefficient clusters identified by the VAE model from the GALFA-\ion{H}{1} dataset. The images are initialized to Gaussian random noise. The top, middle, and bottom rows correspond to the clusters that describe the CNM, WNM, and noise respectively.}
\label{fig:vae_synthsis_galfa}
\end{figure*}

\begin{figure*}[t]
\centering
\includegraphics[width=0.88\textwidth]{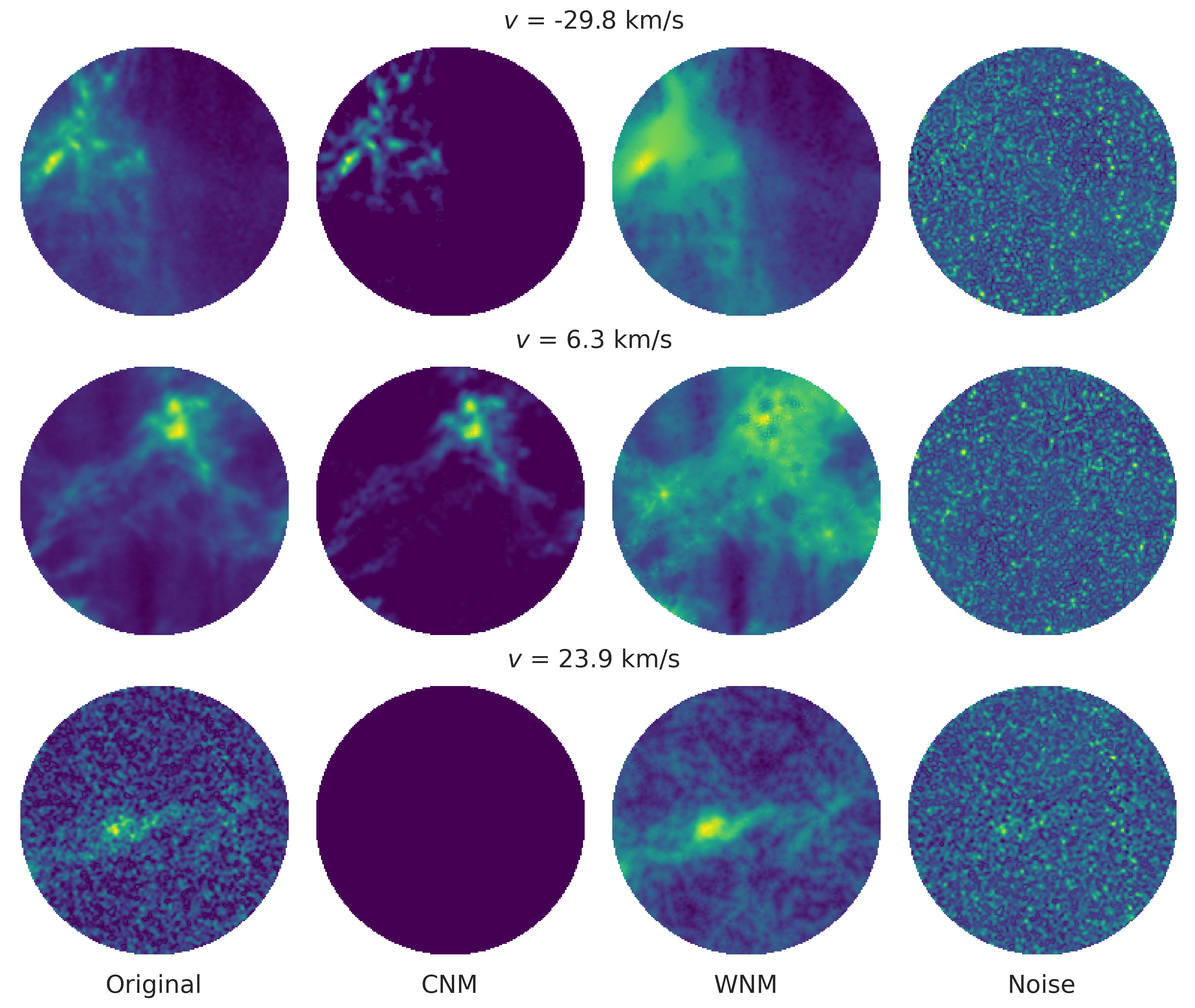}
\caption{Result of SS+VAE phase separation for a example sightline centered at $(\rm{RA=137\degree, Dec=14\degree})$ across different velocity channels. Each column from left to right corresponds to the original image, and the decomposed CNM, WNM, and noise components respectively. The decomposed components show consistent morphological features with what we expect for each phase.}
\label{fig:phase_decomp_galfa}
\end{figure*}

The first step of GALFA-\ion{H}{1} phase separation is identifying distinct clusters from the morphology representation of the dataset using the Gaussian mixture VAE model described in Section \ref{subsec:vae}. Motivated by results from the simulation study in the last section, we choose the number of VAE components to be $c=4$ initially and examine the result of the VAE clustering. Similar to the study with the simulation, we identify three distinct clusters, with an additional cluster overlapping significantly with other components. Unlike the simulation results, the redundant fourth component only overlaps partially with the third component, but both appear noise-like when sampled to produce synthetic realizations (Equation \ref{eq:loss_syn}). This likely indicates that the fourth component encodes partial intra-component variation of the GALFA-\ion{H}{1} noise field, which is more variable than the simple Gaussian noise added to the simulation. We train the VAE model with both 3 and 4 components with qualitatively similar results, but find that $c=4$ leads to better representation of the CNM-like cluster. This is likely due to the imbalanced nature of the GALFA-\ion{H}{1} dataset, which is WNM-dominated generally in the diffuse footprint, and noise-dominated in low-SNR regions at high velocity channels. Having an additional cluster that encodes the intra-class variation of the dominant component helps the latent space better reflect the imbalanced distribution of the input dataset, and better represent the variation of the minority component. Here we present the result of training with $c=4$ components. Future iterations of the model could explore techniques such as resampling and synthetic data generation to directly address imbalanced datasets in VAE training \citep{stocksieker24-ve}. 

To examine if the identified VAE clusters accurately describe the distinct morphological features of the ISM phases, we follow the same process as in the simulation result section. Here, we do not have ground-truth phase information to compare against, so the discussion relies on both physical expectation and comparison with the simulation results. More quantitative comparison will be presented in the following sections when we compare our phase separation results against previous methods. First, we show the distribution of wavelet spectra summary statistics described by Equations \ref{eq:power}-\ref{eq:linearity} for the distinct VAE clusters 0-2 in Figure \ref{fig:prior_dist_galfa}. As in the simulation version in Figure \ref{fig:prior_vs_true_plasmoid}, the distribution and scale-dependent behavior of the three distinct clusters are qualitatively consistent with describing the CNM, WNM, and noise components respectively. Cluster 0 describes a CNM-like component that is sparse and filamentary, particularly towards intermediate and smaller scales, while cluster 2 corresponds to a WNM-like component that is more diffuse and has a wavelet power spectrum with a steeper slope. Cluster 3 behaves similarly to the Gaussian noise added to the simulation dataset, except that it has a larger measure of linearity at small scale $j_1$ and small aspect ratio $\Delta j=|j_2-j_1|$. This could be due to contribution from artifacts such as gain variation and baseline ripple \citep{Peek2017-ql}. 

Image realizations synthesized from VAE cluster samples show a consistent picture. Following the synthesis procedure described by Equation \ref{eq:loss_syn}, where Gaussian noise initialization are evolved to optimize agreement with SS coefficients sampled from one of the VAE clusters, we can generate new image realizations that match the scale-dependent statistical properties of CNM, WNM and noise respectively. We show a few examples for each cluster in Figure \ref{fig:vae_synthsis_galfa}.


\begin{figure*}[t]
\centering
\includegraphics[width=0.88\textwidth]{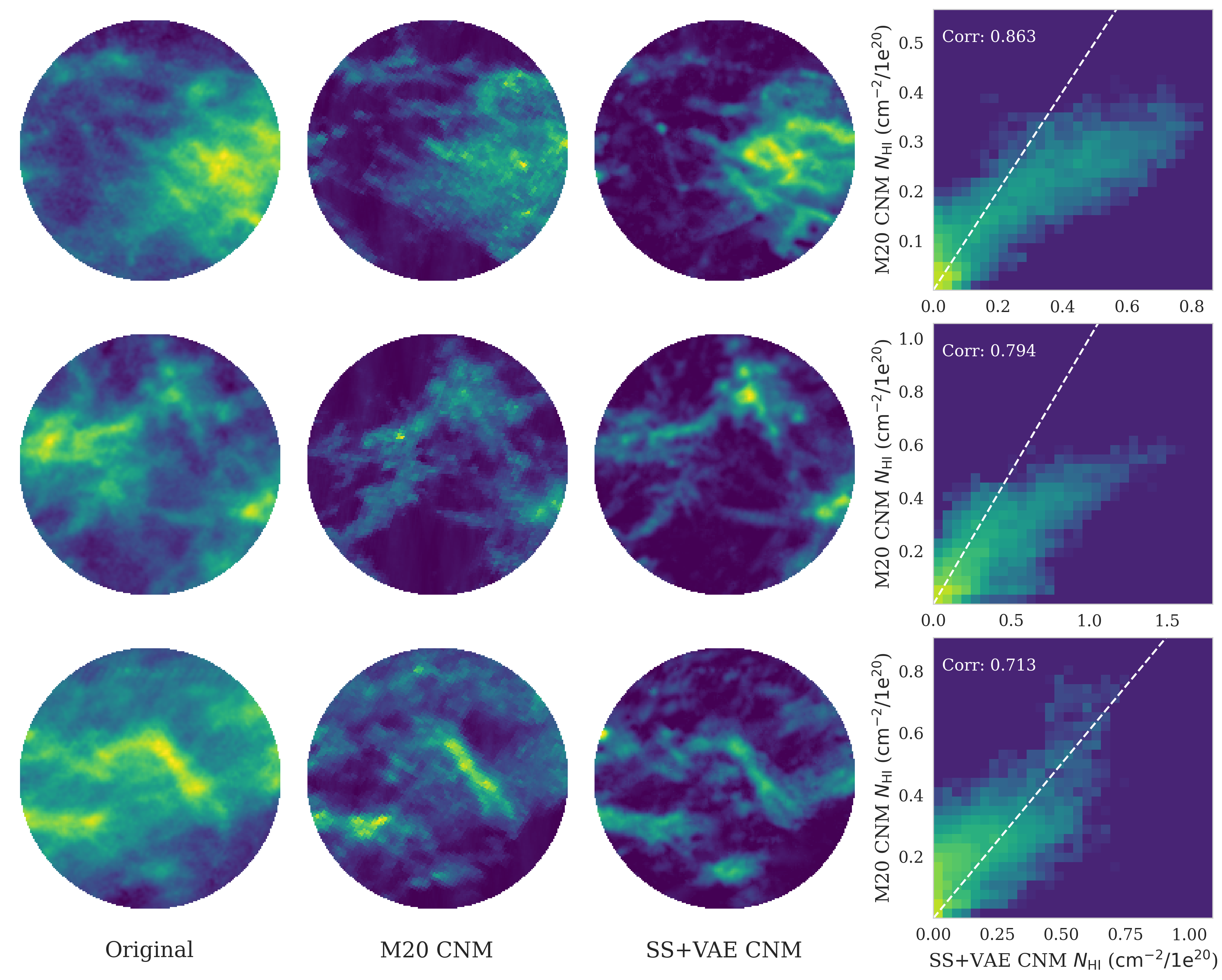}
\caption{Comparing the LOS-integrated CNM column density prediction by the SS+VAE model vs. the M20 map for select regions. From left to right are the original image, the M20 CNM, the SS+VAE CNM, and the correlation between the two CNM maps. The white line indicates one-to-one correspondence. }
\label{fig:m20_vs_wst_patches}
\end{figure*}

\begin{figure*}[t]
\includegraphics[width=0.99\textwidth]{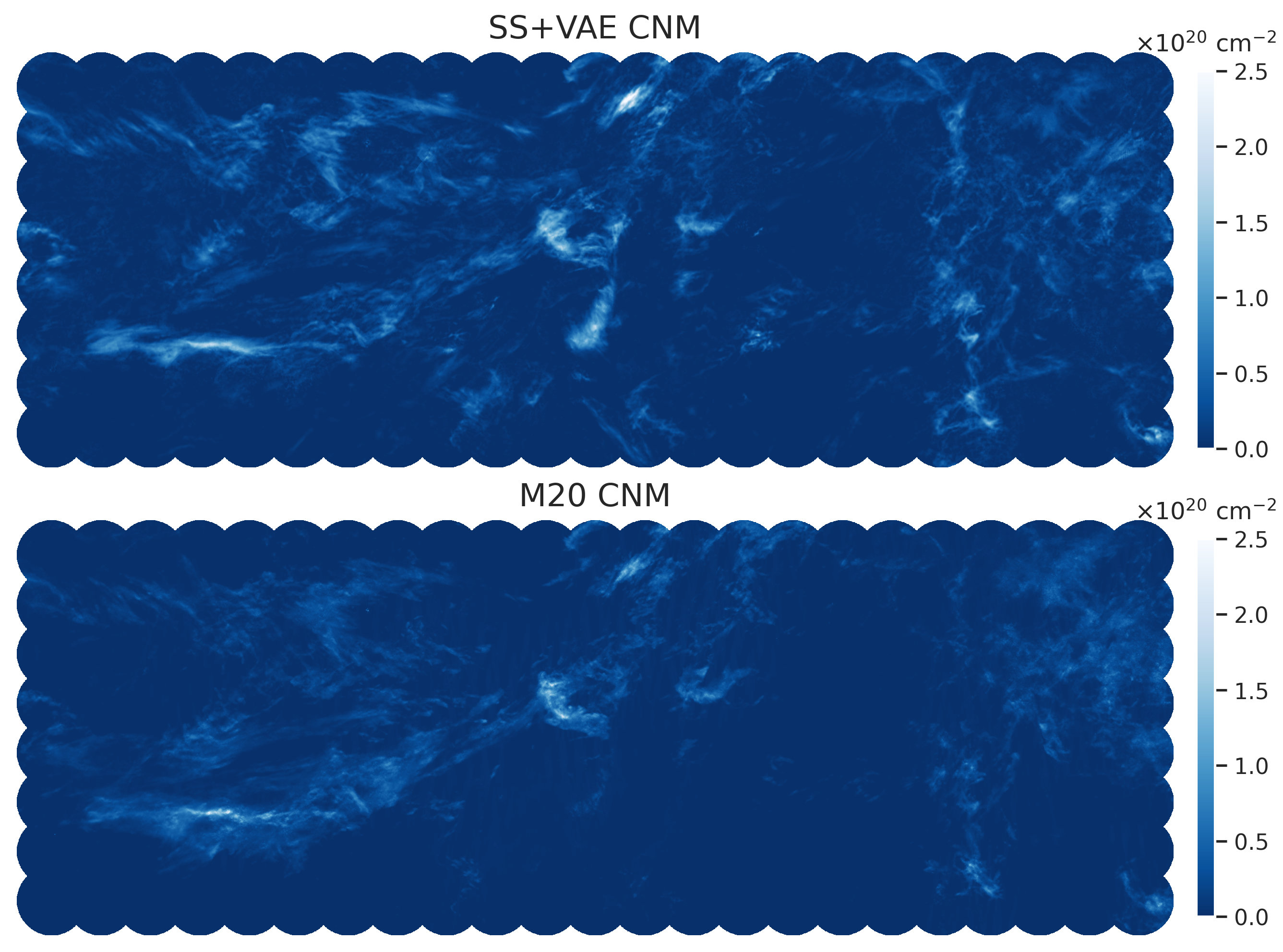}
\caption{Comparing the LOS-integrated CNM column density prediction by the SS+VAE model vs. the M20 map for a continuous $19\degree\times 51\degree$ footprint between $152.3\degree<\mathrm{RA}<203.5\degree$ and $7.8\degree<\mathrm{Dec}<27.0\degree$. The SS+VAE CNM map is stitched together from $256\arcmin \times 256\arcmin$ patches applying weighted average in overlapping regions. The SS+VAE map shows continuous large-scale features and is high correlated with the M20 map with a Pearson coefficient of 0.76 in the footprint shown.}
\label{fig:m20_vs_wst_continuous}
\end{figure*}

\begin{figure*}[t]
\includegraphics[width=0.99\textwidth]{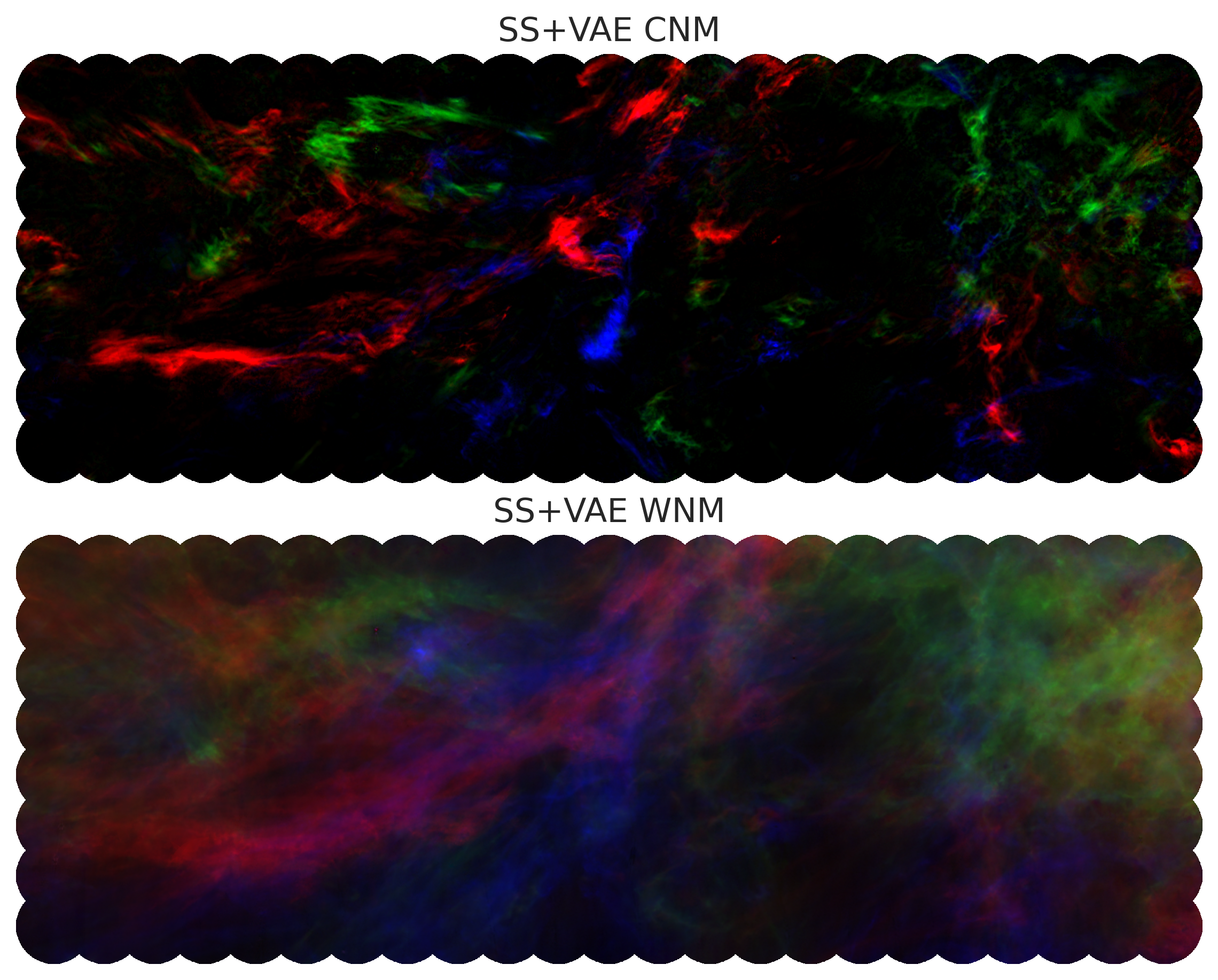}
\caption{The SS+VAE CNM and WNM maps over the same continuous $19\degree\times 51\degree$ footprint as in Figure \ref{fig:m20_vs_wst_continuous}. The red, green, and blue colors correspond to velocity channels between $-3<v<40$ km/s, $-15<v<-3$ km/s, and $-40<v<-15$ km/s respectively.}
\label{fig:wst_continuous_ppv}
\end{figure*}

\begin{figure*}[t]
\includegraphics[width=0.99\textwidth]{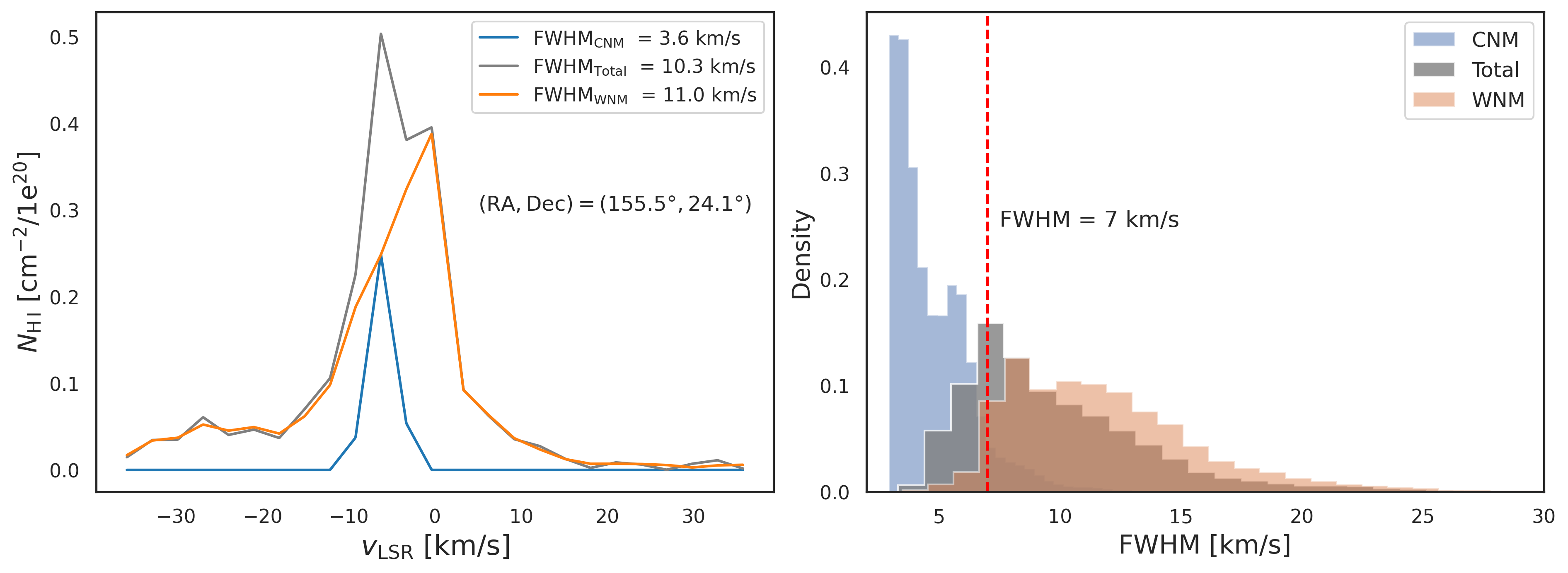}
\caption{Spectral distribution of the total \ion{H}{1} emission and the SS+VAE decomposed CNM and WNM components for sightlines with a minimal level of predicted CNM content of $10^{19}\ \mathrm{cm^{-2}}$.  Left: Spectral distribution of the total, CNM, WNM components for a sample sightline at (RA, Dec)=$(155.5\degree,24.1\degree)$. Right: the histogram of the FWHM distribution for total, CNM, and WNM emission over all sightlines. The red dashed line marks the typical limit of FWHM $< 7$ km/s adopted in Gaussian decomposition of CNM components. 92\% of CNM sightlines have FWHM $< 7$ km/s while 95\% of WNM sightlines have FWHM $> 7$ km/s.}
\label{fig:sepctral_dist}
\end{figure*}

\subsubsection{Decompose \ion{H}{1} Emission with VAE Priors} \label{subsubsec:phase_sep_galfa}

We propose that the VAE clusters shown in Figures \ref{fig:prior_dist_galfa} correspond to different phases in this regime and utilize them as prior information to constrain the decomposition. Sampling from the VAE cluster distribution and making use of the generative process described by Equation \ref{eq:loss_syn}, we produce 300 realizations for each of the CNM, WNM, and noise components respectively, similar to the examples shown in Figure \ref{fig:vae_synthsis_galfa}. The component separation proceeds in two steps. First we perform denoising on the input patches using the noise priors and assuming statistical independence between noise and the data components. Then WNM is separated from the denoised image leaving the residual as the CNM. The CNM is initialized as zero, so no assumptions are introduced except for the morphological constraints learned by the VAE model from the distribution of the input maps. We show the component separation result for an example sightline over multiple velocity channels in Figure \ref{fig:phase_decomp_galfa}. Visual inspection of the different components shows that the phase separation algorithm is able to coherently extract the dense, filamentary CNM-like features from the diffuse WNM-like background for a diverse range of input patches. The denoising also performs well even for very low-SNR patches. However, the noise map separated from some low SNR inputs still show small-scale data residuals, like the third row panels in Figure \ref{fig:phase_decomp_galfa}, which has SNR $< 0.2$. This is related to an limitation of our statistics-based component separation model, which is not deterministic at all scales, and less sensitive to small-scale artifacts. The limitations are discussed more in Section \ref{subsubsec:limitations}, while the denoising results are examined in more detail in Appendix \ref{appx:denoising}. We examine the decomposition of the CNM component more quantitatively in the following section by comparing our CNM map to previous work.

\subsubsection{Comparison with the M20 CNM Map} \label{subsubsec:compare_w_cnn}

We repeat the phase separation process described in the last section on more patches across the high latitude GALFA-\ion{H}{1} sky in 3 km/s velocity channels between $|v|<40$ km/s, and compare the velocity-integrated CNM patches with the CNM map produced using the spectrum-based method from \citet{Murray2020-uf}. The M20 model does not perform full phase separation, but instead predicts a single value for what fraction of each sightline is CNM. We integrate our phase-decomposed CNM patches across velocity channels and compare with the M20 CNM map. In Figure \ref{fig:m20_vs_wst_patches}, we show the comparison in individual $256\arcmin \times256\arcmin$ regions. The SS+VAE CNM patches are generally well-correlated with the M20 patches over the same region, with a Pearson correlation coefficient between $0.7-0.8$. Moreover, the SS+VAE maps seem to recover more coherent structure at small scales, while the M20 maps appear more blocky, due to treating each sightline independently and not utilizing morphological information.

In Figure \ref{fig:m20_vs_wst_continuous}, we show the result of combining 184 overlapping $256\arcmin \times256\arcmin$ SS+VAE CNM patches into one continuous $19\degree\times51\degree$ footprint, compared to the M20 CNM map over the same region. The footprint spans $152.3\degree<\mathrm{RA}<203.5\degree$ and $7.8\degree<\mathrm{Dec}<27.0\degree$ in equatorial coordinates. The discrete SS+VAE patches are stitched together in overlapping regions by a simple weighted average between overlapping patches. The weighting is determined by the apodization mask applied to each patch so that pixels near the center of a patch are weighted more than pixels at the edge of a patch. As Figure \ref{fig:m20_vs_wst_continuous} shows, despite treating each $256\arcmin \times256\arcmin$ region independently and stitching together patches via simple weighted average, the SS+VAE map shows coherent large-scale structures, and is highly correlated with the M20 CNM map, with a Pearson correlation coefficient of 0.76. In comparison, the correlation between the total \ion{H}{1} column density and the M20 CNM column density over this region is 0.59. 

Since the SS+VAE method characterizes and performs component separation on \ion{H}{1} emission patches over individual velocity channels, we go beyond the velocity-integrated CNM predictions in M20, and make phase-decomposed maps in PPV space. In Figure \ref{fig:wst_continuous_ppv}, we show both the SS+VAE CNM and WNM maps over the same $15\degree\times38\degree$ footprint as an RGB image, with each color channel corresponding to integrated column density between velocity channels $-40<v<-15$ km/s, $-15<v<-3$ km/s, and $-3<v<40$ km/s respectively. Both maps show coherent large-scale structures and morphological features consistent with expectations for a sparse, filamentary CNM vs. a diffuse, volume-filling WNM. We discuss the spectral distribution of the decomposed CNM and WNM in more detail in Section \ref{subsubsec:spectra_dist}.

Ultimately, the M20 and SS+VAE maps provide different approaches to extracting CNM information from \ion{H}{1} emission data alone in the absence of absorption information, and come with their own limitations and biases. One limitation of the supervised learning approach adopted by M20 is the potential differences between the simulation training set and the observed data. The authors noted that the M20 map is likely underestimating the CNM fraction due to the low level of CNM in the \citet{kim13-si} simulation that the model is trained on. \citet{marchal24-ft} compared the M20 CNM fraction predictions with their estimate of CNM fraction lower limits derived using Fourier transforms of the emission spectra. They find that while the two values are generally in good agreement, there are many regions at very high latitude ($|b|>60\degree$) where the M20 predictions are lower than their lower limits, particularly towards localized clouds. As Figure \ref{fig:m20_vs_wst_patches} shows, the SS+VAE CNM column density is generally higher than the M20 CNM column. Furthermore, over the $19\degree\times51\degree$ footprint shown in Figure \ref{fig:wst_continuous_ppv}, which covers $59\degree<b<87\degree$, the CNM mass fraction prediction of our model is higher than that of M20 by an average of 0.012. This points to the promising possibility that the SS+VAE model does not suffer from the same problem due to the unsupervised nature of the method -- the phase separating information is learned directly from data without relying on simulation training. The comparison needs to be further validated in the future with the increasing availability of absorption measurements \citep{dickey13-ab, ska15-ab}, but it points to the advantage of having complementary approaches to a challenging, under-constrained problem like ISM phase separation. 

\subsubsection{Spectral Properties of Separated Components} \label{subsubsec:spectra_dist}

While our proposed phase separation method is based purely on morphology information, the fact that it is being applied to regions across individual narrow velocity channels means that we can examine the spectrum of the separated components. Due to the 3 km/s velocity channel width employed in this analysis, the spectral resolution is limited. However, we can still obtain a rough estimate of the line width of each component from the peaks of their spectral distribution and compare that to the expectation for different phases. We do that for all sightlines with at least a minimal level of predicted CNM content of $10^{19}\ \mathrm{cm^{-2}}$ when integrated over the whole sightline, resulting in a total of $\sim 600000$ such spectra. We show the histogram of the FWHM distribution over all sightlines as well as a sample sightline in Figure \ref{fig:sepctral_dist}.  As the histogram shows, the total \ion{H}{1} emission is clearly separated into a narrower CNM component and a broader WNM component. The median FWHM of the CNM component is 4.3 km/s, while the 90th percentile is 6.7 km/s. This WNM component has a median FWHM of 11.2 km/s, and a 90th percentile of 7.6 km/s. The separation of the CNM and WNM components is comparable to the typical limit of FWHM $< 7$ km/s (equivalent to line width $\sigma<3$ km/s) adopted in Gaussian decomposition of CNM components \citep{takakubo66-nh, Murray18-fw, kalberla18-po, marchal19-rs, marchal24-ft}. This result can potentially be refined even further in future work by applying the SS+VAE method to \ion{H}{1} emission channels maps with narrower velocity separation. The consistent spectral behavior of our phase separation result despite no spectral information being used is another indication that the SS+VAE model is extracting relevant phase information from just the morphology of \ion{H}{1} emission patches. 

\section{Discussion} \label{sec:discuss}
By combining morphology characterization with the SS statistics, and unsupervised clustering with a Gaussian-mixture VAE, we present a new morphology-based component separation method capable of extracting phase information directly from \ion{H}{1} emission data. Here we discuss the potential applications of the method in the context of existing literature, as well as limitations and future directions. 

\subsection{Generative Model of \ion{H}{1} Phases and Implications for Galactic Foregrounds} \label{subsubsec:app_phase_gen}
Utilizing SS statistics and Gaussian-mixture VAE model, we built realistic statistical models of different \ion{H}{1} phases that we used to generate prior information for phase separation. More generally, the SS+VAE model outputs provide a generative framework for producing new realizations of CNM, WNM, and noise fields with realistic non-Gaussian features across scales. The process is prescribed by Equation \ref{eq:loss_syn} and examples of generated fields are shown in Figure \ref{fig:prior_dist_galfa}. The framework is entirely data-driven: learned from GALFA-\ion{H}{1} observations without relying on any prior phenomenological assumptions. Realistic statistical descriptions and realizations of different \ion{H}{1} phases can have wide-ranging applications beyond phase separation. One potential example is the construction of Galactic foreground emission models for cosmological studies \citep{pysm25-fg}. \ion{H}{1} emission is a particularly useful tracer of dust reddening and polarization foreground in the context of cosmic microwave background (CMB) observations \citep{lenz17-ln, clark19-fg, bicep23-fg}. \ion{H}{1} gas is strongly coupled to dust emission in the diffuse sky most relevant to CMB studies \citep{lenz17-ln}, and is free from the strong extragalactic contamination that can bias cosmological inference \citep{chiang19-cm}. As a result, some recent studies have utilized \ion{H}{1} emission maps to build dust emission/extinction templates \citep{planck14-hi, lenz17-ln}. However, there is increasing evidence that the degree of dust-to-gas emission correlation is dependent on the mixture of phases along the LOS, due to the variation of dust in different phases \citep{lenz17-ln,nguyen18-dg, clark19_pn, Murray2020-uf}. Therefore, phase-separated maps can be used to make more accurate maps of foreground reddening. Additionally, current Galactic emission data does not provide constraints for all regions and angular scales relevant for CMB foreground studies. Galactic emission models capable of producing mock sky realizations with non-Gaussian ISM features \citep[e.g.][]{pysm25-fg} are important tools for forecasting and data analysis utilized by many current and upcoming CMB experiments \citep[e.g.,][]{simons22-gm, cmbs422-gm, ccat23-gm, spider25-gm, simons25-gm}. Statistical modeling and realization of \ion{H}{1} phases in diffuse regions that capture realistic non-Gaussian features across scale can be important for future efforts of improving Galactic foreground modeling. 

\subsection{\ion{H}{1} Phase Separation in 3D PPV Space} \label{subsubsec:app_ppv_decomp}
In this work we applied the SS+VAE model to separate CNM, WNM+UNM, and noise components from GALFA-HI \ion{H}{1} emission fields across velocity channels $|v|<40$ km/s with channel width $\Delta v=3$ km/s. Most existing non-parametric spectrum-based methods \citep{Murray2020-uf, marchal24-ft, nguyen25-uf} extract a single measure of CNM mass fraction from every sightline without resolving the LOS variation along different velocity channels. Here, we demonstrate that resolving the phase information encoded in \ion{H}{1} emission morphology might be the key to reliably mapping \ion{H}{1} phases in 3D PPV space. In Section \ref{subsec:phase_sep_res}, we show that utilizing only morphology measures of \ion{H}{1} emission patches in narrow velocity channels produces CNM maps that correlate well with the spectrum-based M20 CNM map when integrated over the LOS, while recovering more coherent spatial structures at small scales. In short, the SS+VAE method represents the first \ion{H}{1} phase separation approach to utilize \ion{H}{1} morphology information. The data-driven aspect of the SS+VAE method also enables us to infer phase information directly from observational data, while previous work utilizing supervised learning models requires training on simulations. As a result, the proposed SS+VAE approach complements existing phase separation methods in several important ways, and unlocks a new dimension to resolving the challenging problem of decomposing \ion{H}{1} phases from 21cm emission data alone. 

Large-area, spatially resolved maps of \ion{H}{1} phases are useful for the study of a wide range of problems in ISM science, such as the relationship between \ion{H}{1} phase and dust population \citep{hensley22-pa}, the origin of anomalous microwave emission\ citep{hensley22-pa, ysard22-am}, and the structure of magnetic fields between phases \citep{campbell22-bc, lei24-bc}. The development of high-resolution 3D mapping of dust extinction \citep[e.g.][]{zhang24-dm, edenhofer24-dm} also opens up the exciting prospect of combining 3D PPV phase-decomposed \ion{H}{1} maps with 3D position-position-distance (PPD) dust maps to derive the full kinematic and spatial structure of the neutral ISM. 

\subsection{Current Limitations and Future Directions} \label{subsubsec:limitations}

Ultimately, the analysis here demonstrates that a data-driven, morphology-based approach can be used to characterize \ion{H}{1} phases without utilizing spectral information or relying on simulation data. A complete phase separation method would ideally make use of both spatial and spectral information to decompose \ion{H}{1} emission cubes in 3D PPV space. This is the first step towards that goal, and a similarly data-driven spectral+spatial model utilizing the same SS+VAE technique is in preparation. A limitation of the current spatial-information-only phase separation approach is that the \ion{H}{1} phase realizations and phase-separated maps are susceptible to small-scale artifacts. The constraints on different components in the prior synthesis and component separation process derive solely from the morphology statistics. There is generally not a unique configuration that satisfies the constraints at the pixel level, which means that the resulting phase-separated maps are initial-condition dependent. In other words, small perturbations to the component-separated images can be introduced without significantly changing their SS statistics. Our model is thus a statistical component separation method that leads to results with the right morphology properties at different scales, but does not necessarily recover deterministic structures at all scales, as discussed in other work utilizing ST statistics for component separation \citep{blancard21-dn, delouis22-st, auclair24-st}. This does not significantly affect the large-scale correlation of the phase-separated maps as demonstrated in Section \ref{subsubsec:compare_w_cnn}, but it does mean that individual pixel values are less reliable, particularly in low SNR regions near the GALFA-\ion{H}{1} sensitivity level of $\sim 8\times10^{18} \ \rm{cm^{-2}}$ per 3 km/s channel. 

Another limitation of the current data-driven, morphology-based approach is that our model does not explicitly take into account opacity effects. While most of the high latitude diffuse sky is in the optically thin regime, this assumption breaks in denser regions with prominent CNM features like the North Celestial Pole Loop \citep{tannk22-nc}. Supervised, spectral-based methods like M20 take this account by including optical depth information in their synthetic training data, and find that the ratio between column density computed in the optically thick vs. thin regime increases for regions with higher CNM mass fraction \citep{Murray2020-uf}. With only spatial morphology of emission patches as input, there is no natural way for our current SS+VAE model to recover opacity information. As a result, while our model allows us to decompose brightness temperature observations into contribution from different phases, converting them into column density require optical depth correction that is beyond the scope of the current work.

A follow-up study applying a 1D version of the SS+VAE model that utilizes the 1D scattering transform to characterize \ion{H}{1} spectra is currently in preparation. Comparing these morphology-based phase separation results with the planned spectrum-based work will allow us to characterize any potential orthogonal information between the spatial and spectral dimensions, all within the same data-driven statistical framework. Additionally, the results from the 1D spectrum-based SS+VAE clustering can be used as priors for morphology-based phase separation to provide additional constraints. Similarly, constraints from other spectral-based method such as the $f_{\rm CNM}$ lower limits from \citet{marchal24-ft} can be adopted as priors and compared with the 1D SS+VAE model results. Alternatively, we can explore utilizing 3D wavelet transforms to characterize \ion{H}{1} emission cubes directly in PPV space. There has already been significant progress made towards building fast 3D wavelet transform algorithms \citep{price24-st}. 

\section{Conclusions} \label{sec:conclusion}

In this work, we propose a new data-driven approach utilizing only morphology information in \ion{H}{1} emission channel maps, and demonstrate that it can generate realistic statistical models of \ion{H}{1} phases, which we use as informed priors to produce phase-decomposed CNM, WNM, and noise maps in 3D PPV space in the high-latitude GALFA footprint. Here is a summary of our main results:

\begin{itemize}
    \item We combine SS statistics with a Gaussian-mixture VAE model to identify distinct components directly from \ion{H}{1} emission data. The result is a set of statistical representations that describe the multi-scale morphological features of CNM, WNM, and noise components of \ion{H}{1} emission. We then construct generative models conditioned on these component-separated statistical distributions to produce new realizations of CNM, WNM, and noise fields with realistic multi-scale non-Gaussian features. We utilize these \ion{H}{1} phase realizations as informed priors to decompose the input \ion{H}{1} emission fields. Unlike previous spectrum-based \ion{H}{1} phase separation approaches, here we utilize only \ion{H}{1} emission morphology information without any spectral dimension input, and adopt a data-driven process without relying on simulation training. 
    \item The SS+VAE method is validated by testing it on a high-resolution multiphase ISM simulation from \citet{fielding23}. We demonstrate that the multi-scale morphological features of the components identified by the VAE model from the synthetic total \ion{H}{1} emission agree well with the features of the true CNM, UNM+WNM, and noise components. We find that using the VAE components as priors to decompose synthetic \ion{H}{1} emission patch in $\Delta v=1$ km/s channels between $|v|<20$ km/s recovers the true CNM components in the column density regime that corresponds to the high-latitude diffuse ISM. 
    \item We apply the same data-driven, morphology-based process to observed \ion{H}{1} emission data from the GALFA-\ion{H}{1} survey. The multi-scale morphological features of the distinct VAE components are consistent with a sparse, filamentary CNM, a diffuse, more extended WNM, and a noise component. We produce from the VAE clusters a set of CNM, WNM, and noise realizations via a generative process, and utilize them to decompose the input \ion{H}{1} emission data into $4\degree\times4\degree$ CNM, WNM, and noise patches over 25 velocity channels between $-40\ \rm{km/s}<v<40\ \rm{km/s}$ with channel width $\Delta v=3$ km/s. We combine 184 such overlapping $4\degree\times4\degree$ patches into one continuous $19\degree\times51\degree$ footprint that spans $152.3\degree<\mathrm{RA}<203.5\degree$ and $7.8\degree<\mathrm{Dec}<27.0\degree$.
    \item The decomposed, velocity-integrated CNM patches are compared with the CNM maps produced from the spectrum-based CNN model from \citet{Murray2020-uf} (M20). We find that the morphology-based CNM maps are highly correlated with the spectrum-based maps with a Pearson correlation coefficient of $\sim 0.8$, while recovering more spatially coherent structures at small scales. 
    \item Decomposing \ion{H}{1} emission patches in PPV space across velocity channels also allows us to examine the spectral distribution of the CNM and WNM components. Despite incorporating no spectral information in the component separation process and a coarse sampling of velocity channels at $\Delta v=3$ km/s, we find that the 90th percentile of the spectral FWHM of the CNM is 6.7 km/s, while the 10th percentile of the WNM FWHM is 7.6 km/s. The separation between the CNM and WNM spectral width is comparable to the typical limit of FWHM $< 7$ km/s adopted in Gaussian decomposition of CNM components.
    \item These results demonstrate the potential of the proposed SS+VAE method as a general approach to tackle under-constrained component separation problems in a data-driven way. In the \ion{H}{1} phase separation context, the performance of the morphology-based method shows that there is significant phase-sensitive information in the spatial dimension of \ion{H}{1} emission, untapped by previous phase-decomposition methods. Follow-up work currently in preparation will extend the SS+VAE method to combine both 1D spectral + 2D spatial information to decompose 3D \ion{H}{1} emission data cubes under the same data-driven framework. 
\end{itemize}

We will make the SS+VAE algorithm\footnote{\url{https://github.com/minjielei/stvae_ism}} discussed in this work, in addition to the phase-separated GALFA-\ion{H}{1} maps\footnote{\url{https://doi.org/10.7910/DVN/EXJLF3}}, publicly available upon publication.

\vspace*{5mm}

\section{Acknowledgments}
We thank the anonymous reviewer for thoughtful comments that improved this work. We are grateful to Claire Murray and Douglas Finkbeiner for helpful discussions. This work was supported by the National Science Foundation under grants No. AST-2106607 and AST-2441452 (PI S.E.C.). S.E.C. additionally acknowledges support from an Alfred P. Sloan Research Fellowship. I.S.B was supported by NASA through the Hubble Fellowship, grant HST-HF2-51525.001-A awarded by the Space Telescope Science Institute, which is operated by the Association of Universities for Research in Astronomy, Incorporated, under NASA contract NAS 5-26555.

This work benefited from the conference ``Structure and polarization in the interstellar medium: A Conference in Honor of Prof. John Dickey", a hybrid meeting hosted jointly at Stanford University and at the Australia Telescope National Facility in February 2025. We acknowledge support from the National Science Foundation (NSF Award No. 2502957), from the Kavli Institute for Particle Astrophysics and Cosmology, from the Commonwealth Scientific and Industrial Research Organisation, and from the Australian Research Council. The computations in this paper were run on the Sherlock cluster, supported by the Stanford Research Computing Center at Stanford University.

\vspace{5mm}

\software{astropy \citep{astropy13}, 
          scattering \citep{cheng2021-si},
          NumPy \citep{numpy11},
          healpy \citep{healpy05-hp},
          PyTorch \citep{pytorch19}
          }

\begin{figure*}[t]
\includegraphics[width=0.99\textwidth]{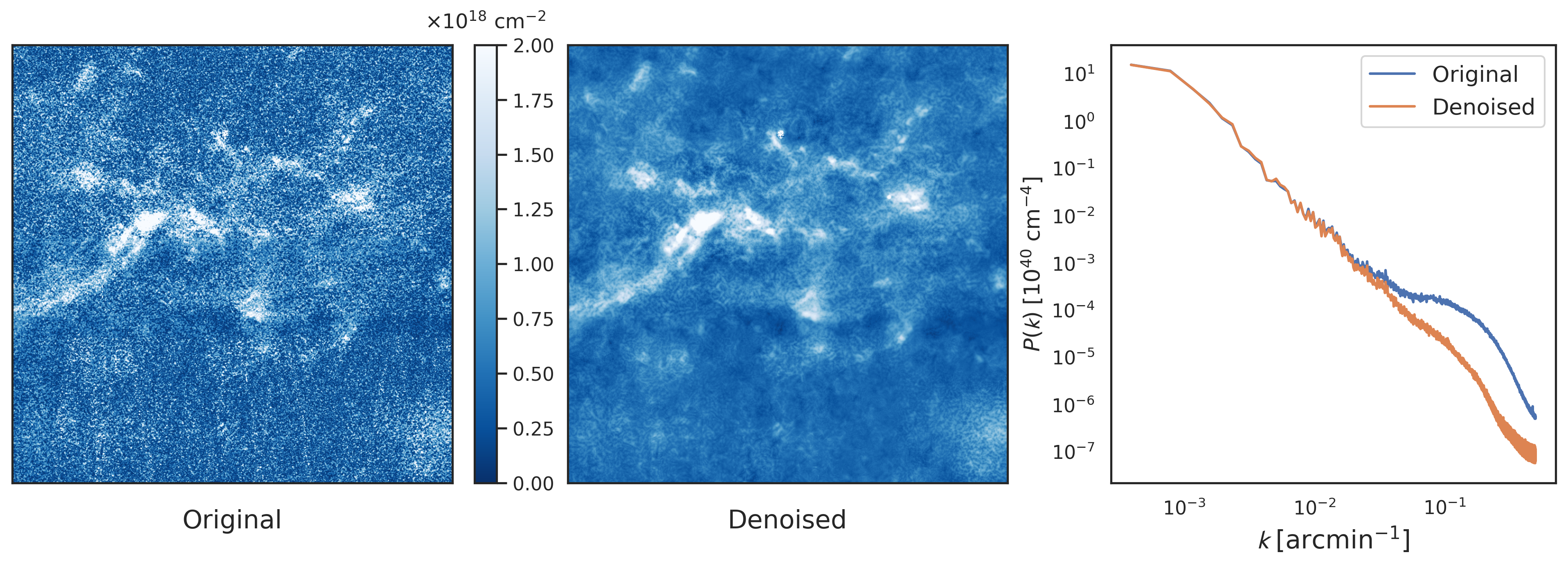}
\caption{GALFA-\ion{H}{1} denoising result shown for a $17\degree\times17\degree$ region at $v=18$ km/s, spanning $185.3\degree<\mathrm{RA}<202.4\degree$ and $8.9\degree<\mathrm{Dec}<26.0\degree$. The left and middle panels show the original GALFA-\ion{H}{1} emission map and the SS+VAE denoised map, respectively. The right panel compares the power spectra of the original and denoised maps. }
\label{fig:galfa_denoise}
\end{figure*}

\appendix

\section{Gaussian-Mixture VAE Architecture and Training Details} \label{appx:training}

We adopt the Gaussian-mixture VAE implementation described in \citet{siahkoohi24-va}. The encoder is composed of a series of four fully connected residual blocks. The first fully connected layer reduces the dimensionality of the input features to a hidden dimension chosen to be 1024, while the last layer further reduces the hidden dimension to a latent dimension set of 32. Each fully connected layer is followed by a Batchnorm layer and LeakyReLU nonlinearity. The decoder mirrors the encoder architecture and reconstructs the input feature space from the latent space. 

To train the Gaussian-mixture VAE model, we employ the Adam optimization algorithm \citep{kingma14-ad} with a learning rate of $10^{-3}$.  We run the training for 300 epochs with a batch size of 512. To enable differentiable approximate sampling for categorical variables, we utilize the Gumbel-Softmax distribution with initial temperature parameter of 1.0 \citep{Jang16-gs}. We then exponentially decay the temperature to a minimum value of 0.7, with a decay parameter of $\ln(2)/50$ so that the temperature halves every 50 epochs. During training, we randomly reserve 10\% of the data as validation. The VAE training was performed on a Tesla V100 and takes approximately 20 minutes. The subsequent phase separation step is performed on the same device and takes $\sim6$ minutes to decompose eight $4\degree\times4\degree$ patches in parallel.

\section{GALFA-\ion{H}{1} Denoising Results} \label{appx:denoising}

While most existing non-parametric phase separation methods is trained or designed to specifically predict the CNM fraction for a given sightline \citep{murray18-sp, marchal24-ft, nguyen25-uf}, we adopt an agnostic, data-driven approach that allows us to identify the noise as well as CNM and WNM components directly from the input data, and produce a denoised map as part of the component separation process. In Figure \ref{fig:galfa_denoise}, we show the denoising result for a $17\degree\times17\degree$ GALFA-\ion{H}{1} region with low SNR of $\sim1$ at $v=18$ km/s. Visual comparison between the original and denoised map show that the SS+VAE algorithm reduces the noise while preserving small-scale data structure. The prominent basket-weave artifacts in the bottom part of the image are also significantly reduced. Comparing the power spectra of the denoised and original map, the denoised map spectrum is better approximated by a single power law, and is consistent with the original map at large scales, while containing less power at small scales where we expect noise to dominate. 


\bibliography{stvae4hi}{}
\bibliographystyle{aasjournal}



\end{document}